\documentclass[a4paper,fleqn,usenatbib]{mnras}

\usepackage{newtxtext,newtxmath}
\usepackage[T1]{fontenc}
\usepackage{ae,aecompl}
\usepackage{times}
\usepackage{enumitem}
\usepackage{times}
\usepackage{textcomp}
\usepackage{verbatim}
\usepackage{esdiff}

\newcommand{\msun}{{\,\rm M_\odot}}
\newcommand{\Lsun}{{\,\rm L_\odot}}
\newcommand{\kms}{\,{\rm km}\,{\rm s}^{-1}}

\newcommand{\Myr}{\,{\rm Myr}}

\newcommand{\pkpc}{\,{\rm pkpc}}
\newcommand{\Mpc}{\,{\rm Mpc}}
\newcommand{\pMpc}{\,{\rm pMpc}}

\newcommand{\mmag}{\,{\rm mag}}

\def\aap{A\&A}
\def\apj{ApJ}

\def\apjl{ApJ}
\def\mnras{MNRAS}
\def\araa{ARA\&A}
\def\aj{AJ}

\def\physrep{Phys. Rep.}
\def\nat{Nature}

\def\apjs{ApJS}

\usepackage{graphicx}	% Including figure files
\usepackage{amsmath}	% Advanced maths commands
\makeatletter
\newcommand{\rmnum}[1]{\romannumeral #1}
\newcommand{\Rmnum}[1]{\expandafter\@slowromancap\romannumeral #1@}

\renewcommand\paragraph{\@startsection{paragraph}{4}{\z@}{3.25ex\@plus1ex\@minus.2ex}{-1em}{\normalfont\it\normalsize}}

\defcitealias{Vogelsberger2020}{Paper~\Rmnum{1}}
\defcitealias{Shen2020}{Paper~\Rmnum{2}}

\title[High redshift predictions from IllustrisTNG]
{High redshift predictions from IllustrisTNG: \Rmnum{3}. Infrared luminosity functions, obscured star formation and dust temperature of high-redshift galaxies}

      \author[Shen et al.] {\parbox{18.0cm}{
          Xuejian Shen$^1$\thanks{email:xshen@caltech.edu},
          Mark Vogelsberger$^2$,
          Dylan Nelson$^3$, 
          Sandro Tacchella$^4$, 
          Lars Hernquist$^4$,
          Volker Springel$^5$,
          Federico Marinacci$^6$,
          Paul Torrey$^7$
        }\vspace{0.3cm}\\
        $^1$ TAPIR, California Institute of Technology, Pasadena, CA 91125, USA\\
        $^2$ Department of Physics, Kavli Institute for Astrophysics and Space Research, Massachusetts Institute of Technology, Cambridge, MA 02139, USA\\
        $^3$ Universit\"{a}t Heidelberg, Zentrum f\"{u}r Astronomie, Institut f\"{u}r theoretische Astrophysik, Albert-Ueberle-Str. 2, 69120 Heidelberg, Germany\\
        $^4$ Center for Astrophysics $\vert$ Harvard \& Smithsonian, 60 Garden Street, Cambridge, MA, 02138, USA\\
        $^5$ Max-Planck-Institut f\"{u}r Astrophysik, Karl-Schwarzschild-Str. 1, D-85741 Garching, Germany\\
        $^6$ Department of Physics \& Astronomy ``Augusto Righi'', University of Bologna, via Gobetti 93/2, I-40129 Bologna, Italy\\
        $^7$ Department of Astronomy, University of Florida, 211 Bryant Space Sciences Center, Gainesville, FL 32611, USA
}

\pubyear{2022}

\begin{document}

\label{firstpage}
\pagerange{\pageref{firstpage}--\pageref{lastpage}}
\maketitle

\begin{abstract} 
We post-process galaxies in the IllustrisTNG simulations with {\sc Skirt} radiative transfer calculations to make predictions for the rest-frame near-infrared (NIR) and far-infrared (FIR) properties of galaxies at $z\geq 4$. The rest-frame $K$- and $z$-band galaxy luminosity functions from TNG are overall consistent with observations, despite $\sim 0.5\,\mathrm{dex}$ underprediction at $z=4$ for $M_{\rm K}\lesssim -25$ and $M_{\rm z}\lesssim -24$. Predictions for the {\it JWST} MIRI observed galaxy luminosity functions and number counts are given. Based on theoretical estimations, we show that the next-generation survey conducted by {\it JWST} can detect 500 (30) galaxies in F1000W in a survey area of $500\,{\rm arcmin}^{2}$ at $z=6$ ($z=8$). As opposed to the consistency in the UV, optical and NIR, we find that TNG, combined with our dust modelling choices, significantly underpredicts the abundance of most dust-obscured and thus most luminous FIR galaxies. As a result, the obscured cosmic star formation rate density (SFRD) and the SFRD contributed by optical/NIR dark objects are underpredicted. The discrepancies discovered here could provide new constraints on the sub-grid feedback models, or the dust contents, of simulations. Meanwhile, although the TNG predicted dust temperature and its relations with IR luminosity and redshift are qualitatively consistent with observations, the peak dust temperature of $z\geq 6$ galaxies are overestimated by about $20\,{\rm K}$. This could be related to the limited mass resolution of our simulations to fully resolve the porosity of the interstellar medium (or specifically its dust content) at these redshifts.
\end{abstract}


%%%%%%%%%%%%%%%%%%%%%%%%%%%%%%%%%%%%%%%%%%%%%%%%%%

\pubyear{2021}

%%%%%%%%%%%%%%%%% BODY OF PAPER %%%%%%%%%%%%%%%%%%

\section{Introduction}

The $\Lambda{\rm CDM}$ model~\citep[e.g.,][]{Planck2016, Planck2020} is the standard theoretical paradigm for structure formation. In this framework, initial small density perturbations grow via gravitational instability and produce the large-scale structures as well as the bound dark matter haloes where galaxies form. Based on this, the theory of galaxy formation~\citep[e.g.,][]{White1978,Blumenthal1984,Cole2000} makes predictions that can be tested by observed galaxy populations. Aside from the well-studied constraints in the local Universe, galaxies formed in the early Universe provide a new testing ground for galaxy formation theories~\citep[see reviews of][and references therein]{Shapley2011,Stark2016,Dayal2018}, with open questions related to star formation in dense molecular clouds and stellar/supernovae feedback, the metal enrichment and dust formation in early environments, the seeding of massive black holes, the triggering of AGN activity, etc.

Due to the limited wavelength coverage of the Hubble Space Telescope ({\it HST}) and the insufficient sensitivity of infrared (IR) instruments, the observation of high-redshift galaxies was mainly performed in the rest-frame ultra-violet \citep[UV; e.g.,][]{Bouwens2003,Wilkins2010,McLure2013,Finkelstein2015,Oesch2018,Bouwens2019}. However, UV observations are inadequate for revealing the entire galaxy population. It is known that the cosmic star formation rate density at low redshift is dominated by dusty star-forming galaxies~\citep[DSFGs;][]{Magnelli2011,Casey2012,Gruppioni2013} that are heavily obscured in optical and UV while bright at far-infrared (FIR) wavelengths. At high redshift, owing to instrumental limitations, the abundance of such galaxies and their contribution to cosmic star formation are still highly uncertain~\citep[e.g.,][]{Casey2018}. In recent years, ALMA has been identifying some FIR-bright but UV-faint galaxies at $z\gtrsim 3$~\citep[e.g.,][]{Simpson2014,Wang2019,Yamaguchi2019,Franco2020,Dud2020} and measuring the dust continuum emission from these galaxies. These observations reveal galaxies that were hidden in previous optical/NIR selections and may be the tip-of-the-iceberg of the highly obscured high-redshift galaxy population. Several future sub-millimeter/millimeter instruments have been proposed, including the TolTEC camera on the Large Millimeter Telescope \citep[LMT;][]{Bryan2018}, the Origins Space Telescope \citep[OST;][]{Battersby2018} and the Chajnantor Sub-Millimeter Survey Telescope \citep[CSST;][]{Golwala2018}, which will help to reveal the demographics of DSFGs at high redshift. 

Meanwhile, the James Webb Space Telescope ({\it JWST}; \citealt{Gardner2006}) will be in operation soon. As discussed in \citet[][\citetalias{Vogelsberger2020} of this series]{Vogelsberger2020}, the NIRCam of {\it JWST} will push the detection of galaxies in the UV to the fainter end and help reveal lower mass galaxies, which have a significant contribution to the cosmic SFRD, as well as more heavily obscured DSFGs. In addition, the NIRCam and the Mid-Infrared Instrument (MIRI) will provide photometric and spectroscopic access to the rest-frame UV to mid-infrared (mid-IR) spectral energy distributions (SEDs) of galaxies at high redshift. Rest-frame optical and NIR observations would be particularly useful for an unbiased measurement of galaxy stellar mass and constraints on the star formation histories of galaxies. Mid-IR observations of dust emission lines would shed light on the physical properties of dust in high-redshift galaxies. Undoubtedly, the next-generation galaxy surveys will provide a much deeper and broader spectral coverage of galaxy emission at high redshift, particularly at IR wavelengths. The advancement will hopefully provide a more complete picture of galaxy formation in the early Universe, aside from the UV emission from the unobscured young stellar populations.

In parallel to the observational efforts, theoretical predictions are also necessary for the study of high-redshift galaxies. Several attempts have been made with semi-analytical models of galaxy formation \citep[e.g.,][]{Clay2015,Liu2016,Cowley2018,Tacchella2018,Yung2018,Yung2019} paired with simple prescriptions for dust extinction, either empirical dust corrections based on observational relations or simple dust shell models that link extinction with integrated optical depth. Similar prescriptions have been adopted in cosmological simulations \citep[e.g.,][]{Cullen2017,Wilkins2017,Ma2018}. However, such treatments are great simplifications of the scattering and absorption processes of dust grains and effectively neglect the complicated dust geometry. The predictive power for the dust continuum emission at IR wavelengths is also limited. Alternatively, radiative transfer calculations have been introduced for many galaxy simulations to model dust (or neutral gas) absorption and emission. For example, \citet{Cen2014} ran radiative transfer calculations for a sample of $198$ galaxies in cosmological zoom-in simulations at $z\sim 7$, with predictions for IR properties of ALMA galaxies; \citet{Camps2016,Trayford2017} performed radiative transfer calculations on the EAGLE simulation \citep{Crain2015,Schaye2015} and made predictions for UV-to-sub-millimeter SEDs of galaxies in the local Universe; \citet{Ma2019} applied radiative transfer calculations to the FIRE-2 \citep{Hopkins2018} simulations and studied dust extinction and emission in $z\geq 5$ galaxies. Similar techniques have also been used to study the physical origin and variations of the IRX-$\beta$ relation and dust attenuation curves \citep[e.g.,][]{Safarzadeh2017, Narayanan2018,Liang2020,Schulz2020}. Several radiation-hydrodynamical simulations (with on-the-fly radiative transfer calculations) \citep[e.g,][]{Rosdahl2013,Kimm2014,Ocvirk2016,Kimm2017,Rosdahl2018} have been performed to study cosmic reionization on large scales and the escape of ionizing photons from early galaxies.

The IllustrisTNG project is a series of large, cosmological magneto-hydrodynamical simulations of galaxy formation \citep{Nelson2019b,Pillepich2019}. The TNG simulations have been calibrated and tested by numerous low-redshift observables \citep[e.g.,][]{Springel2018,Pillepich2018b,Nelson2018,Naiman2018,Marinacci2018,Genel2018} and offer an unprecedented stand point for the study of high-redshift galaxy populations. In \citetalias{Vogelsberger2020}, we developed a radiative transfer post-processing pipeline to calculate SEDs and images of galaxies at $z\geq2$ in the TNG simulations. The pipeline was calibrated based on the UV luminosity functions that are well constrained by observations. In \citetalias{Vogelsberger2020} and \citet[][\citetalias{Shen2020} of the series]{Shen2020}, we made predictions for the rest-frame UV luminosity functions, optical emission line luminosity functions, UV continuum and optical spectral indices. We found good agreement with observations, except for a missing population of heavily obscured, UV red galaxies in the TNG simulations. Here, we aim to adapt the pipeline for predictions for the IR properties of galaxies. The major advantages of TNG compared to other cosmological simulations are its representative box size to allow statistical predictions for galaxy populations (even at high redshift) and its reasonable mass and spatial resolution to describe the multi-phase interstellar medium (ISM), star formation and feedback processes. Unlike some simulations dedicated for high-redshift studies, TNG has been evolved to low redshift and tested against various low-redshift observables.

This paper is organized as follows: In Section~\ref{sec:sim}, we briefly describe the IllustrisTNG simulation suite and state the numerical parameters of TNG50, TNG100 and TNG300 in detail. In Section~\ref{sec:method}, we describe the method we used to derive the dust attenuated broadband photometry and SEDs of galaxies in the simulations. The main results are presented in Section~\ref{sec:results}, where we make various predictions and comparisons with observations. The summary and conclusions are presented in Section~\ref{sec:conclusions}. In Appendix \ref{appsec:skirt_test}, we present additional tests of different {\sc Skirt} configurations and numeric convergence.

\begin{table*}
\begin{tabular}{llccrrrccccc}
\hline
{\bf IllustrisTNG Simulation} & run & {volume side length} & $N_{\rm gas}$  & $N_{\rm dm}$ & $m_{\rm b}$ & $m_{\rm dm}$ & $\epsilon_{\rm dm,stars}$ &  $\epsilon_{\rm gas}^{\rm min}$ \\
 & &   $[h^{-1}{\rm Mpc}]$  & & &  $[h^{-1}{\rm M}_\odot]$ & $[h^{-1}{\rm M}_\odot]$ &  $[h^{-1}{\rm kpc}]$ & $[h^{-1}{\rm kpc}]$\\
\hline
\hline
{\bf TNG300}   &  TNG300(-1)  & 205  & $2500^3$  & $2500^3$   & $7.4\times 10^6$   & $4.0\times 10^7$   & 1.0 & 0.25\\
\hline
{\bf TNG100}   &  TNG100(-1)  & 75   & $1820^3$  & $1820^3$   & $9.4\times 10^5$   & $5.1\times 10^6$   & 0.5 & 0.125 \\
\hline
{\bf TNG50}    &  TNG50(-1)   & 35   & $2160^3$  & $2160^3$   & $5.7\times 10^4$   & $3.1\times 10^5$   & 0.2 & 0.05 \\

\hline
\end{tabular}
\caption{{\bf IllustrisTNG simulation suite.} The table shows the basic numerical parameters of the three primary IllustrisTNG simulations: simulation volume side length, number of gas cells ($N_{\rm gas}$), number of dark matter particles ($N_{\rm dm}$), baryon mass resolution ($m_{\rm b}$), dark matter mass resolution ($m_{\rm DM}$), Plummer-equivalent maximum physical softening length of dark matter and stellar particles ($\epsilon_{\rm dm,stars}$), and the minimal comoving cell softening length $\epsilon_{\rm gas}^{\rm min}$. In the following we will refer to TNG50-1, TNG100-1 and TNG300-1 as TNG50, TNG100 and TNG300, respectively. 
\label{tab:tabsims}}
\end{table*}

\section{Simulation}
\label{sec:sim}
The analysis here is based on the IllustrisTNG simulation suite~\citep[][]{Marinacci2018, Naiman2018, Nelson2018, Pillepich2018b, Springel2018}, including the newest addition, TNG50~\citep[][]{Nelson2019b, Pillepich2019} with the highest numerical resolution in the suite. The IllustrisTNG simulation suite is the follow-up project to the Illustris simulations~\citep{Vogelsberger2014a, Vogelsberger2014b, Genel2014, Nelson2015, Sijacki2015}. The simulations were conducted with the moving-mesh code {\sc Arepo} \citep{Springel2010,Pakmor2016,Weinberger2020} and adopted the IllustrisTNG galaxy formation model~\citep[][]{Weinberger2017,Pillepich2018a}, which is an update of the Illustris galaxy formation model~\citep{Vogelsberger2013, Torrey2014}. The IllustrisTNG simulation suite consists of three primary simulations: TNG50, TNG100 and TNG300 \citep{Nelson2019a}, covering three different periodic, uniformly-sampled volumes, roughly ${\approx 50^3}, 100^3, 300^3\,{\rm Mpc}^3$. For simplicity, in the following, we will refer to TNG50-1, TNG100-1 and TNG300-1 as TNG50, TNG100 and TNG300, respectively. The simulations employ the following cosmological parameters~\citep[][]{Planck2016}: $\Omega_{\rm m} = 0.3089$, $\Omega_{\rm b} = 0.0486$, $\Omega_{\Lambda} = 0.6911$, $H_0 = 100\,h\,\kms \Mpc^{-1} = 67.74\,\kms \Mpc^{-1}$, $\sigma_{8} = 0.8159$, and $n_{\rm s} = 0.9667$. The numerical parameters of the simulations are summarized in Table~\ref{tab:tabsims}. 

\section{Galaxy IR luminosities and SEDs}
\label{sec:method}

In this section, we introduce the approach we adopt to calculate dust-attenuated/intrinsic IR SEDs and band luminosities/magnitudes of galaxies in the IllustrisTNG simulations. We follow the Model C procedure introduced in \citetalias{Vogelsberger2020} and adapt it for IR predictions. In this section, we will briefly review the procedure and refer the readers to \citetalias{Vogelsberger2020} for more details. 

In this work, we define a galaxy as being either a central or satellite galaxy as identified by the {\sc Subfind} algorithm~\citep[][]{Springel2001, Dolag2009}. For all the following analysis, we impose a stellar mass cut for galaxies. We only consider galaxies with a stellar mass larger than $100$ times the baryonic mass resolution, $100 \times m_{\rm b}$, within twice the stellar half mass radius. Galaxies resolved with a lower number of resolution elements will not be considered, since we assume that their structure is not reliably modelled. 

The calculation of galaxy SEDs consists of two steps: {\it \rmnum{1}}) characterize the radiation source and determine the spatial and wavelength distribution of the intrinsic emission; {\it \rmnum{2}}) perform dust radiative transfer calculations, including dust absorption, dust self-absorption and dust emission. In the first step, we assigned intrinsic emission to stellar particles in the simulations according to their ages and metallicities using the stellar population synthesis method. To be specific, we adopt the Flexible Stellar Population Synthesis ({\sc Fsps}) code~\citep{Conroy2009,Conroy2010} to model the intrinsic SEDs of old stellar particles ($t_{\rm age}>10\Myr$) and the {\sc Mappings-\Rmnum{3}} SED library~\citep{Groves2008} to model those of young stellar particles ($t_{\rm age}<10\Myr$). The {\sc Mappings-\Rmnum{3}} SED library self-consistently considers the dust attenuation in the birth clouds of young stars which cannot be properly resolved in the simulations. In the second step, we perform the full Monte Carlo dust radiative transfer calculations using a modified version of the publicly available {\sc Skirt} (version 8)\footnote{\href{http://www.skirt.ugent.be/root/_landing.html}{http://www.skirt.ugent.be/root/\_landing.html}} code~\citep{Baes2011,Camps2013,Camps2015,Saftly2014}. Modifications were made to incorporate the {\sc Fsps} SED templates into {\sc Skirt}. Photon packages are randomly released based on the source distribution characterized by the positions and SEDs of stellar particles. The emitted photon packages will further interact with the dust in the ISM. To determine the distribution of dust in the ISM, we select cold, star-forming gas cells~(with ${\rm SFR}>0$ or temperature $<8000\,{\rm K}$) from the simulations and calculate the metal mass distribution based on the metallicities of selected cells. We assume that dust is traced by metals in the ISM and turn the metal mass distribution into the dust mass distribution with a constant, averaged dust-to-metal ratio of all galaxies at a fixed redshift. To be specific, this dust-to-metal ratio depends on redshift as $0.9 \times (z/2)^{-1.92}$, which has been calibrated based on the observed UV luminosity functions at $z=2\operatorname{-}10$ as introduced in \citetalias{Vogelsberger2020}.

The dust mass distribution is then mapped onto an adaptively refined octree grid for radiative transfer calculations with the refinement criterion chosen to match the spatial resolution of the simulations. Ultimately, after photons fully interact with dust in the galaxy and escape, they are collected by a mock detector $1\pMpc$~\footnote{Physical $\Mpc$.} away from the simulated galaxy along the positive z-direction of the simulation coordinates. The integrated galaxy flux is then recorded, which provides us with the dust-attenuated SED of the galaxy in the rest frame. Galaxy SEDs without the resolved dust attenuation are derived in the same way with no resolved dust distribution included. We note that we use the term ``without the resolved dust'' because the unresolved dust component in the {\sc Mappings-\Rmnum{3}} SED library is always present. Compared with the resolved dust attenuation, the impact of the unresolved dust attenuation on galaxy continuum emission is limited. For rest-frame broadband photometry, galaxy SEDs are convolved with the transmission curves using the {\sc Sedpy}\footnote{\href{https://github.com/bd-j/sedpy}{https://github.com/bd-j/sedpy}} code. For the calculation of apparent band magnitudes, the rest-frame flux is redshifted, corrected for intergalactic medium absorption~\citep{Madau1995,Madau1996} and converted to the observed spectra. In addition to the rest-frame opical/UV bands and NIRCam bands of {\it JWST} studied in \citetalias{Vogelsberger2020}, we add rest-frame NIR bands \citep[J, H, Ks bands of the 2MASS survey][]{2Mass2006} and the MIRI bands of {\it JWST}. 
\begin{figure*}
    \centering
    \includegraphics[width=0.99\textwidth]{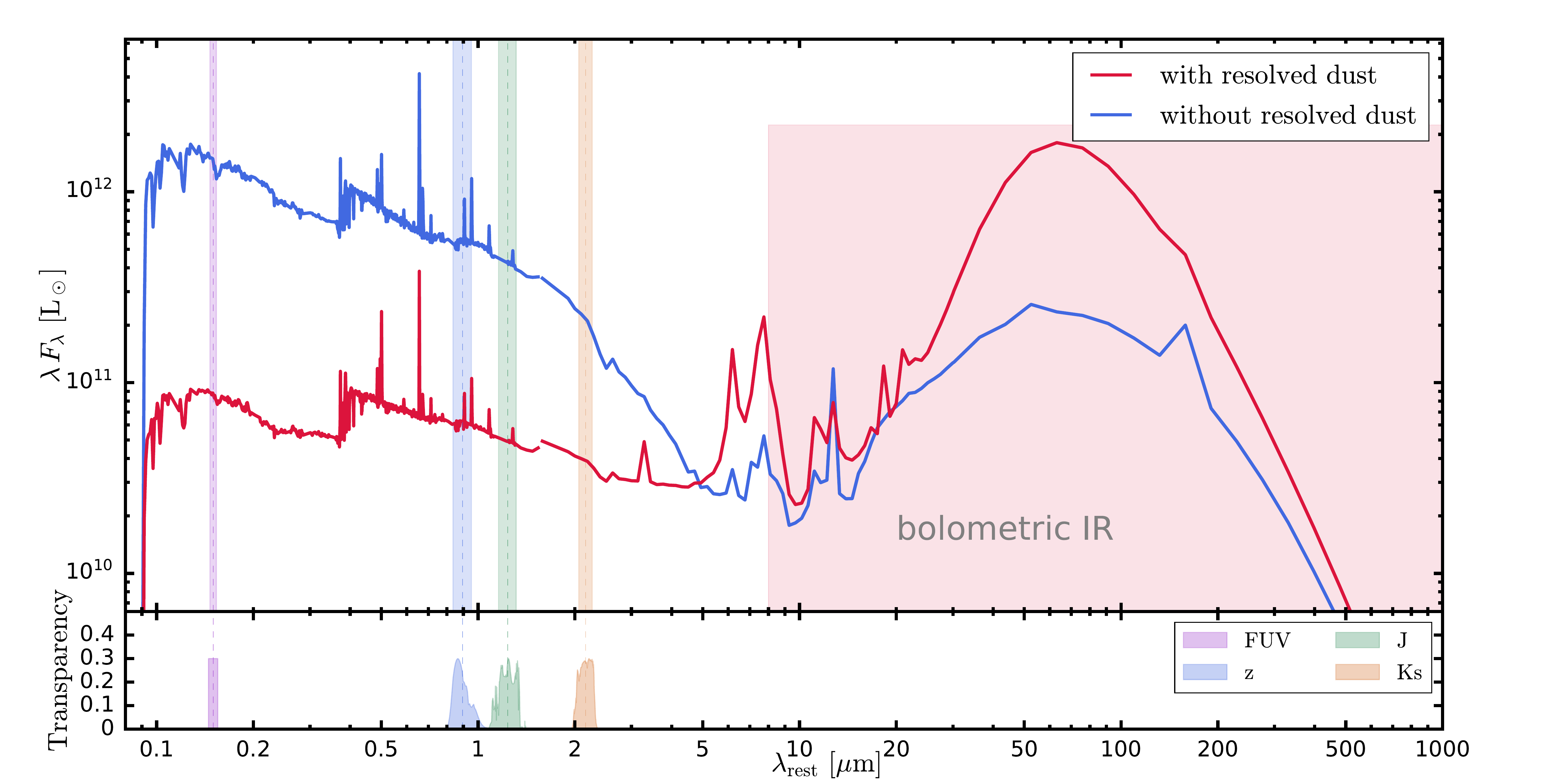}
    \includegraphics[width=0.99\textwidth]{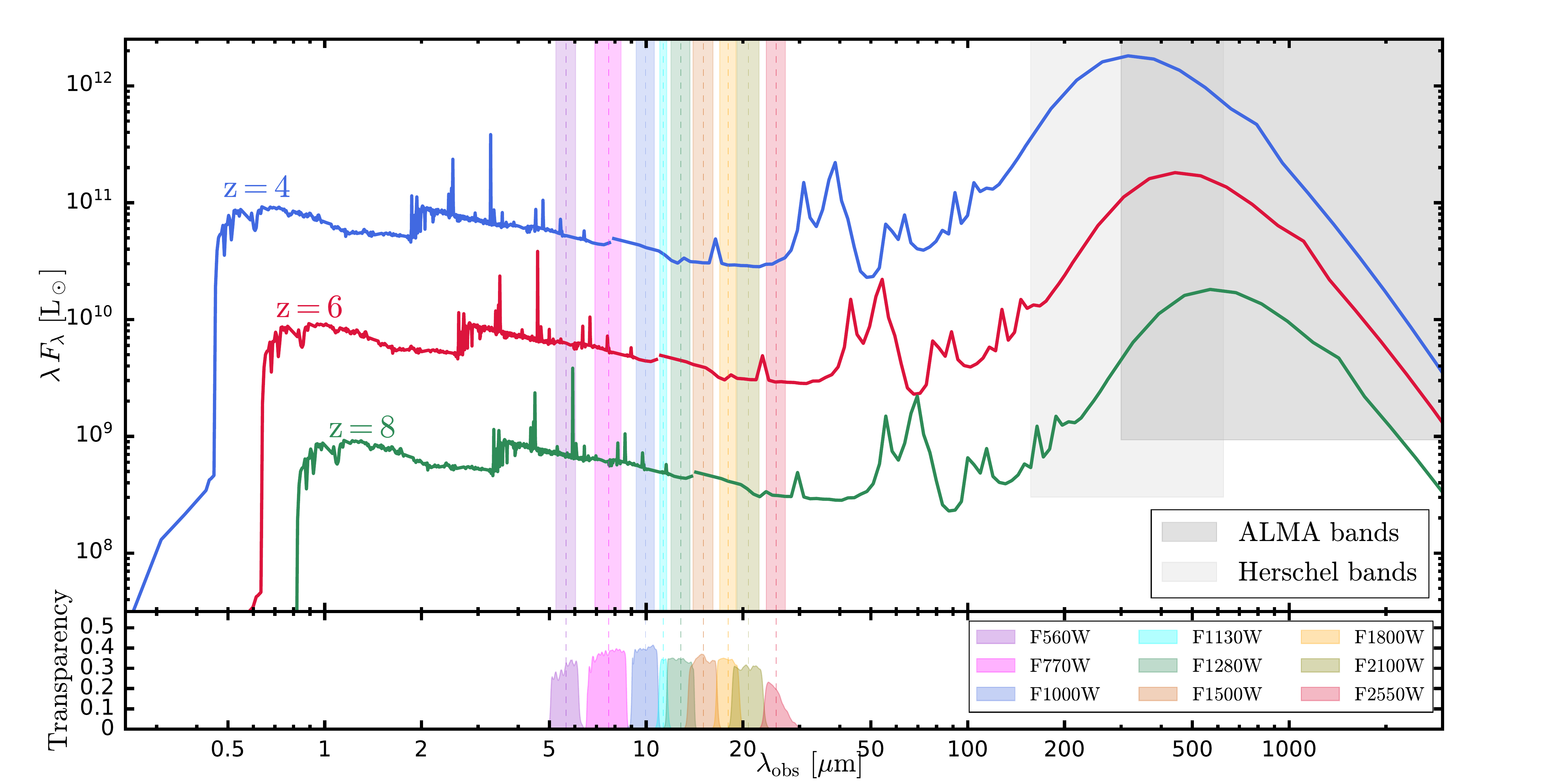}
    \caption{\textbf{Galaxy SEDs in rest-frame or observer's frame.} The SEDs are derived from the radiative transfer post-processing of a star-forming galaxy in TNG100. The UV to optical SED derived in \citetalias{Vogelsberger2020} can be smoothly connected to (without renormalization) the IR SED derived in this work at $\lambda \sim 1.7\micron$. In the top panel, we show the rest-frame SEDs with/without resolved dust attenuation and emission, along with the rest-frame bands involved in this work. The dashed lines show pivot wavelengths and the shaded regions indicate effective bandwidths. The bolometric IR luminosity is defined as the integrated flux in $8\operatorname{-}1000\micron$ as shown by the red shaded region. The lower sub-panel shows the transmission curves of the bands. In the bottom panel, we show the SEDs (all with resolved dust attenuation and emission) in the observer's frame assuming that the galaxy is located at $z=2,\,4,\,6$. The {\it JWST} MIRI broad bands and the wavelength coverage of ALMA and Herschel bands are shown for comparison. The transmission curves of the MIRI bands are also shown in the lower sub-panel.}
    \label{fig:sed_example}
\end{figure*}

In \citetalias{Vogelsberger2020} and \citetalias{Shen2020}, we limited our predictions to rest-frame UV/optical properties of galaxies. To extend the predictions to IR wavelengths, the setup of the radiative transfer calculations has been modified in the following aspects:
\begin{itemize}

    \item The wavelength grid is adapted to have $101$ points with a wider wavelength coverage. We first evenly sample $51$ points from $0.1$ to $1000\micron$ as the base grid. To better resolve the emission lines in mid-IR wavelengths, we create a refined grid in the mid-IR, sampling $61$ points from $2$ to $30\micron$. Combining the base grid and the refined grid in the mid-IR, we get the final grid consisting of $101$ points. Following \citetalias{Vogelsberger2020}, by default, the number of photon packages per wavelength grid is set to be the number of bound stellar particles $N_{\rm star}$ in the galaxy, with $10^{2}$ ($10^5$) as the minimum (maximum) number. In empirical tests, we find that this choice of photon package numbers gives converged galaxy UV-to-FIR SEDs, except for galaxies that are close to the selection criteria ($N_{\rm star}\sim 100$, see the stellar mass cut we adopt above). To account for this, for calculations on TNG50, we increase the number of photon packages and the minimum number by a factor of three to achieve better convergence in low-mass, poorly resolved galaxies. Further increasing the number of photon packages will not lead to differences in galaxy SEDs. Poorly resolved galaxies in TNG100 and TNG300 can be resolution corrected based on well-resolved TNG50 galaxies, if the results show significant differences. Details of the convergence tests and discussion of the choice of photon package numbers are shown in Appendix~\ref{appsec:skirt_test}.
    
    \item Non-local thermal equilibrium is considered. Small dust grains are allowed to be stochastically heated and decouple from local thermal equilibrium. In order to trace grains of different sizes separately, we switch our dust model from the \citet{Draine2007} model to the \citet{Zubko2004} multi-grain model, which has been adopted in \citet{Camps2016,Trayford2017} and \citet{Schulz2020}. Similar to the \citet{Draine2007} model adopted in \citetalias{Vogelsberger2020}, the \citet{Zubko2004} model includes a composition of graphite, silicate and polycyclic aromatic hydrocarbon (PAH) grains. The size distributions and the relative amount of the dust grains are chosen so as to reproduce the dust properties of the Milky Way. Different from the model in \citetalias{Vogelsberger2020} (using the average dust properties of the \citet{Draine2007} dust mixture), the \citet{Zubko2004} model traces dust grains of different sizes separately. We adopt $10$ bins for grain sizes for each type of dust and further increasing the number of bins to $15$ does not lead to differences in photometric predictions. The impact of the dust model on galaxy SEDs is illustrated in Figure~\ref{appfig:sed_skirt_config} in Appendix~\ref{appsec:skirt_test}.
    
    \item In practice, we find that switching to the \citet{Zubko2004} multi-grain model (see above) leads to $\sim 0.2\operatorname{-}0.3 \mmag$ underprediction on the UV luminosities of galaxies. To compensate for that, we decrease the dust-to-metal ratio in the new runs by $25\%$ compared to the calibrated model in \citetalias{Vogelsberger2020}. After this modification, the UV magnitudes of galaxies are consistent with the results of \citetalias{Vogelsberger2020} with $\leq 0.1 \mmag$ differences and the total absorbed luminosities integrated from UV to optical are consistent with \citetalias{Vogelsberger2020} with $\leq 0.1\,\mathrm{dex}$ differences.
    
    \item Dust self-absorption and re-emission are included. The dust emission procedure is carried out iteratively until the total luminosity absorbed by dust converges at a $<3\%$ level. This mainly affects the peak of the FIR continuum and has little impact in the UV, optical and NIR, as shown in Figure~\ref{appfig:sed_skirt_config} in Appendix~\ref{appsec:skirt_test}.
    
    \item The inclusion of dust self-absorption significantly increases the computational cost of the radiative transfer calculations. Therefore, for this work, we limit the radiative transfer calculations to three selected snapshots corresponding to $z=4,\,6,\,8$, rather than the redshift range $z=2\operatorname{-}10$ covered by \citetalias{Vogelsberger2020,Shen2020}. For $z=4$, we perform the calculations for all three simulations. For $z=6$ and $z=8$, we perform the calculations only for TNG100 and TNG300, the dynamical ranges of which are sufficient to match IR observations of galaxies.

\end{itemize}

In the top panel of Figure~\ref{fig:sed_example}, we show integrated SEDs from the radiative transfer post-processing of a star-forming galaxy in TNG100. In the top panel, we show the rest-frame SEDs with/without resolved dust attenuation. The ``resolved'' dust refers to the dust resolved by the simulations and involved in radiative transfer calculations, and the ``unresolved'' dust refers to the one associated with the {\sc Mappings-\Rmnum{3}} SED model. It is encouraging that the UV to optical SED derived in \citetalias{Vogelsberger2020} (with better wavelength resolution and no dust emission included) can be smoothly connected to (without renormalization) the IR SED derived in this work at $\lambda \sim 1.7\micron$. In the figure, the rest-frame bands involved in this work are shown for comparison. The rest-frame FUV band ($1500\text{\AA}$) is sensitive to the young stellar populations and thus the on-going star formation in the galaxy. The rest-frame NIR bands ($z,\,J,\,Ks$) are sensitive to the emission of old stellar populations and thus the integrated star formation in the past. Notably, the luminosities in these bands are less affected by dust attenuation and serve as good indicators for galaxy stellar mass. The bolometric IR luminosity is defined as the integrated flux in the wavelength range $8\operatorname{-}1000~\micron$, which is dominated by dust continuum emission and indicates the total amount of energy absorbed by dust at short wavelengths. In the bottom panel, we show the apparent SEDs (all with resolved dust attenuation) in observer's frame assuming that the galaxy is at $z=4,\,6,\,8$. For reference, we show the {\it JWST} MIRI bands and the wavelength coverage of Herschel and ALMA bands at FIR. The comparison demonstrates MIRI's promise for measuring the rest-frame optical and NIR emission of galaxies at high redshift. Paired with Herschel/ALMA observation at longer wavelengths and NIRCam observation in rest-frame UV, a fairly complete coverage of galaxy rest-frame SED can be achieved.

\section{Results}
\label{sec:results}

\subsection{NIR band luminosity functions}

\begin{figure}
    \centering
    \includegraphics[width=0.49\textwidth]{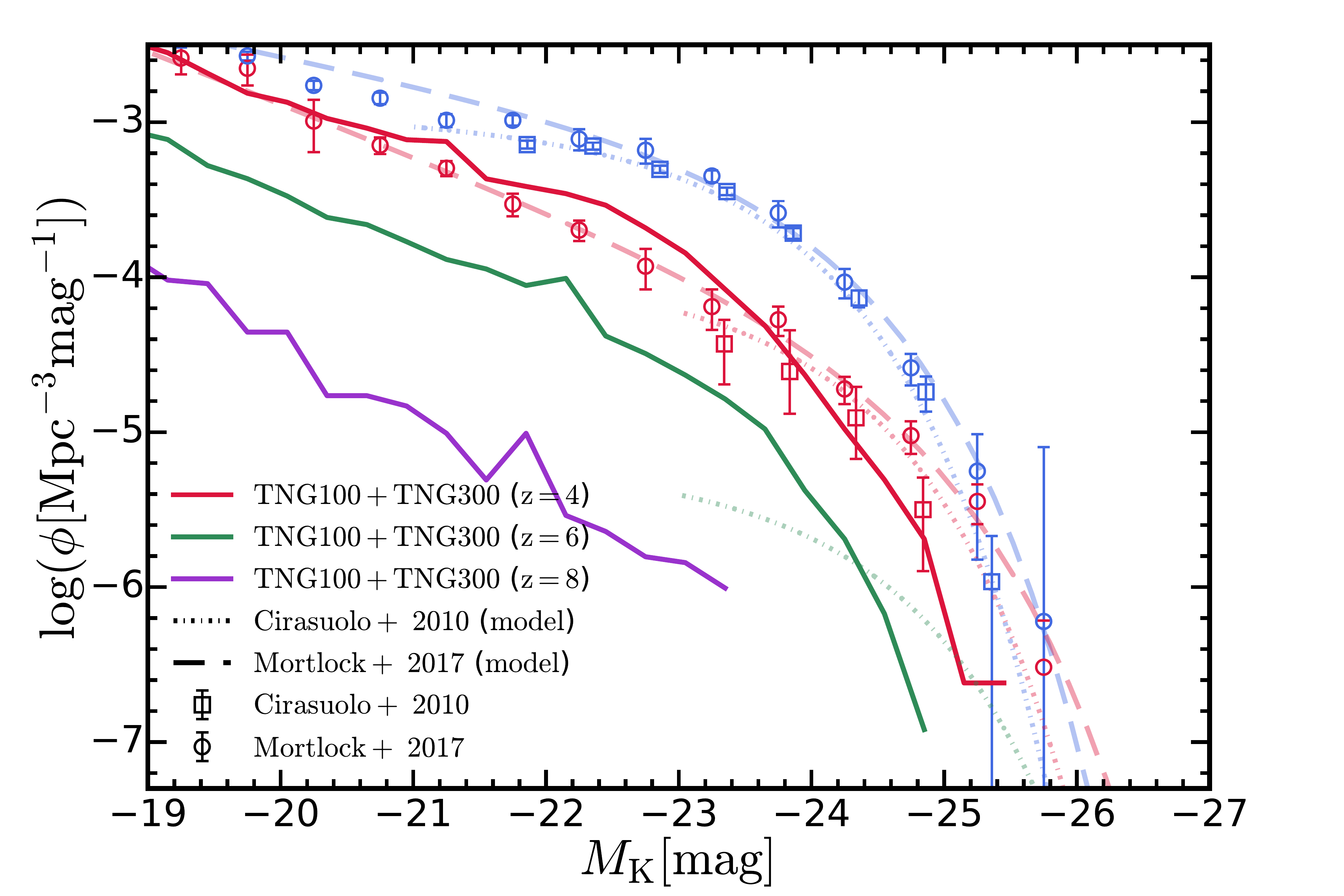}
    \caption{\textbf{Galaxy $K$-band luminosity function.} The galaxy rest-frame $K$-band luminosity functions at $z=4,\,6,\,8$ from the IllustrisTNG simulations are presented in solid lines as labelled. Binned estimations from observations~\citep{Cirasuolo2010,Mortlock2017} are shown with open markers. The best-fit Schechter functions in these work are shown in dashed and dotted lines. Note that the $z=6$ result from \citet{Cirasuolo2010} is based on a model-dependent extrapolation of the observations at $z\lesssim 4$. Observational constraints at $z=2,\,4$ and $6$ are shown in blue, red and green colors, respectively. The predicted $K$-band luminosity functions are consistent with observations, especially the steep faint-end slope as revealed by most recent observations.}
    \label{fig:kband-lf}
\end{figure}

\begin{figure}
    \centering
    \includegraphics[width=0.49\textwidth]{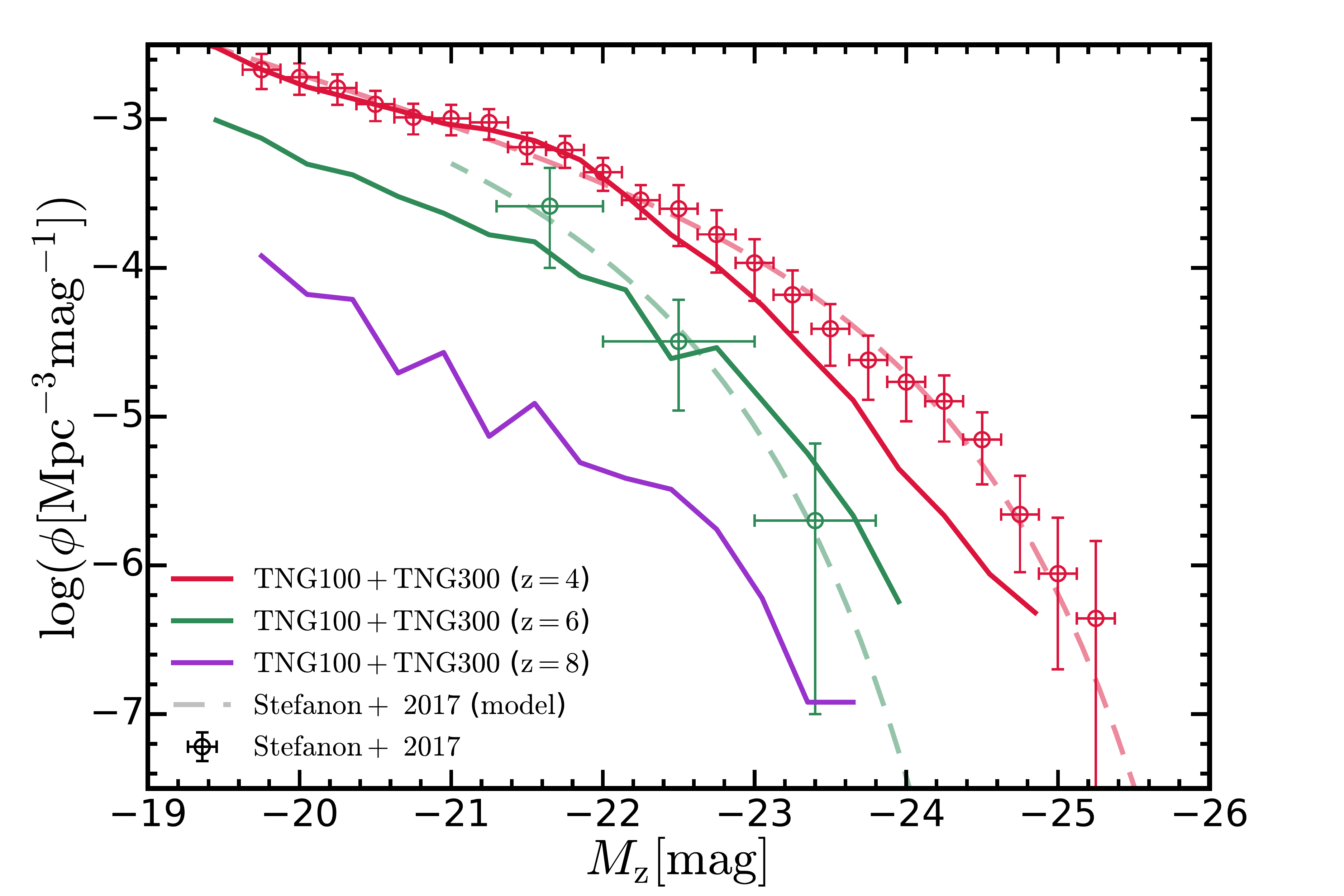}
    \caption{\textbf{Galaxy $z$-band luminosity function.} The galaxy rest-frame $z$-band luminosity functions at $z=4,\,6,\,8$ from the IllustrisTNG simulations are presented in solid lines as labelled. Binned estimations and Schechter fits from observations~\citep{Stefanon2017} are shown with open markers and dashed lines. Observational constraints at $z=4$ and $6$ are shown in red and green colors, respectively. The predicted $z$-band luminosity functions are consistent with observations. The evolutionary pattern is similar to that of the $K$-band luminosity function.}
    \label{fig:zband-lf}
\end{figure}

Rest-frame NIR luminosities of galaxies are sensitive to the old stellar populations and are less affected by dust absorption or emission, on-going star formation or the choice of population synthesis model. Therefore, the NIR luminosity function is an ideal indicator for the stellar mass assembly history of galaxies in the Universe. The rest-frame $K$-band (including the Ks variant) centered around $2.2\micron$ has been widely used for such studies at high redshift~\citep[e.g.,][]{Pozzetti2003,Drory2003,Caputi2006,Saracco2006,Cirasuolo2010,Mortlock2017}. In Figure~\ref{fig:kband-lf}, we present the galaxy rest-frame $K$-band luminosity functions at $z=4,\,6,\,8$ predicted from the IllustrisTNG simulations and compare them with observations~\citep{Cirasuolo2010,Mortlock2017}. The predictions from TNG at different redshifts are shown in solid lines with different colors. Luminosity functions from TNG100 and TNG300 are combined here following the procedure described in \citetalias{Vogelsberger2020}. We note that resolution corrections are not applied here since we find that the rest-frame NIR band luminosity functions from TNG50, TNG100 and TNG300 agree well in their shared dynamical ranges at the redshifts considered. Similar agreement is found for the MIRI apparent band luminosity functions. Binned estimations from observations are shown with open markers. The best-fit Schechter functions in these observational studies are shown in dashed lines (truncated at the observational limit). One set of measurements in \citet{Mortlock2017} was originally performed at $z\simeq 3.25$. To compare it with our predictions at $z=4$, we correct the binned estimations for the decrease in the number density normalization by linearly extrapolating their best-fit $\phi^{\ast}$ in single Schechter fits at $z\leq 3.25$ to $z=4$. The Schechter fit at $z=4$ from \citet{Mortlock2017} is also obtained by linearly extrapolating their best-fit $\phi^{\ast}$ to $z=4$ while keeping $M^{\ast}$ and $\alpha$ the same as those at $z=3.25$. The galaxy $K$-band luminosity function at $z\gtrsim 2$ can be well characterized by a single Schechter function~\citep{Mortlock2017}. Previous measurements of the $K$-band luminosity function at $z\gtrsim 1$~\citep{Caputi2006,Saracco2006,Cirasuolo2010} suggested a shallow faint-end slope of $\alpha\simeq -1$, independent of redshift. However, updated measurements~\citep{Mortlock2017}, which probed much fainter luminosities than previous studies, revealed a steep faint-end slope of $-2 \lesssim \alpha \lesssim -1.5$ at $z \gtrsim 2$ and argued that previous studies were incomplete at faint luminosities. Overall, the TNG predictions agree well with the observational constraints and the faint-end slopes are as steep as suggested by \citet{Mortlock2017}. However, compared to the observational results, the simulation prediction at $z=4$ exhibits a ``bump'' at $M_{\rm K}\sim -22.5\mmag$. Although being consistent with \citet{Cirasuolo2010} results, the simulation prediction at $z=4$ falls below the more recent \citet{Mortlock2017} measurements at $M_{\rm K}\sim -25\mmag$ by about $0.5$ dex. In terms of the redshift evolution of the $K$-band luminosity function at $z\gtrsim 2$, TNG predicts that the number density normalization continuously decreases towards higher redshift while the faint-end slope remains the same. Approaching $z\sim 2$, the evolution of the bright end luminosity function stalls, which is likely related to the quenching of star formation in massive galaxies.

The rest-frame $z$-band centered around $0.9\micron$ is also used in studies of NIR luminosity functions and measurements of galaxy stellar mass functions~\citep[e.g.,][]{Stefanon2017}. It is located at shorter wavelengths than the $K$-band. It can thus be probed to higher redshift with current instruments and it remains free from contamination of nebular emission lines and the Balmer break. In Figure~\ref{fig:zband-lf}, we present the galaxy $z$-band luminosity function at $z=4,\,6,\,8$ from the IllustrisTNG simulations and compare them with observations~\citep{Stefanon2017}. Both the binned estimations and the Schechter fits from the observations are presented. The predicted luminosity functions are overall consistent with observations, despite that the predicted number density at $M_{\rm z}\sim -24$ at $z=4$ is about $0.5\,\mathrm{dex}$ lower than observations. Such discrepancy is not found at $z=6$ or in the $K$-band luminosity functions. The redshift evolution of the $z$-band luminosity function is similar to that of the $K$-band luminosity function presented above. At $z\geq 4$, the number density normalization continuously decreases towards higher redshift while the faint-end slope remains the same. Since passive galaxies have very limited contribution to the NIR luminosity functions (or stellar mass functions) of galaxies at high redshift~\citep[e.g.,][]{Mortlock2017}, the evolutionary pattern of the NIR luminosity functions reflects the mass assembly of star-forming galaxies at high redshift. The weak evolution of the faint-end slope indicates that the logarithmic increase of galaxy luminosities with cosmic time is roughly uniform across luminosities, and the mass growth rates (mass doubling times) of high-redshift galaxies are roughly independent of their masses. 

\subsection{JWST MIRI bands apparent luminosity functions}
\label{sec:results-jwstlf}

\begin{figure*}
    \centering
    \includegraphics[width=0.49\textwidth]{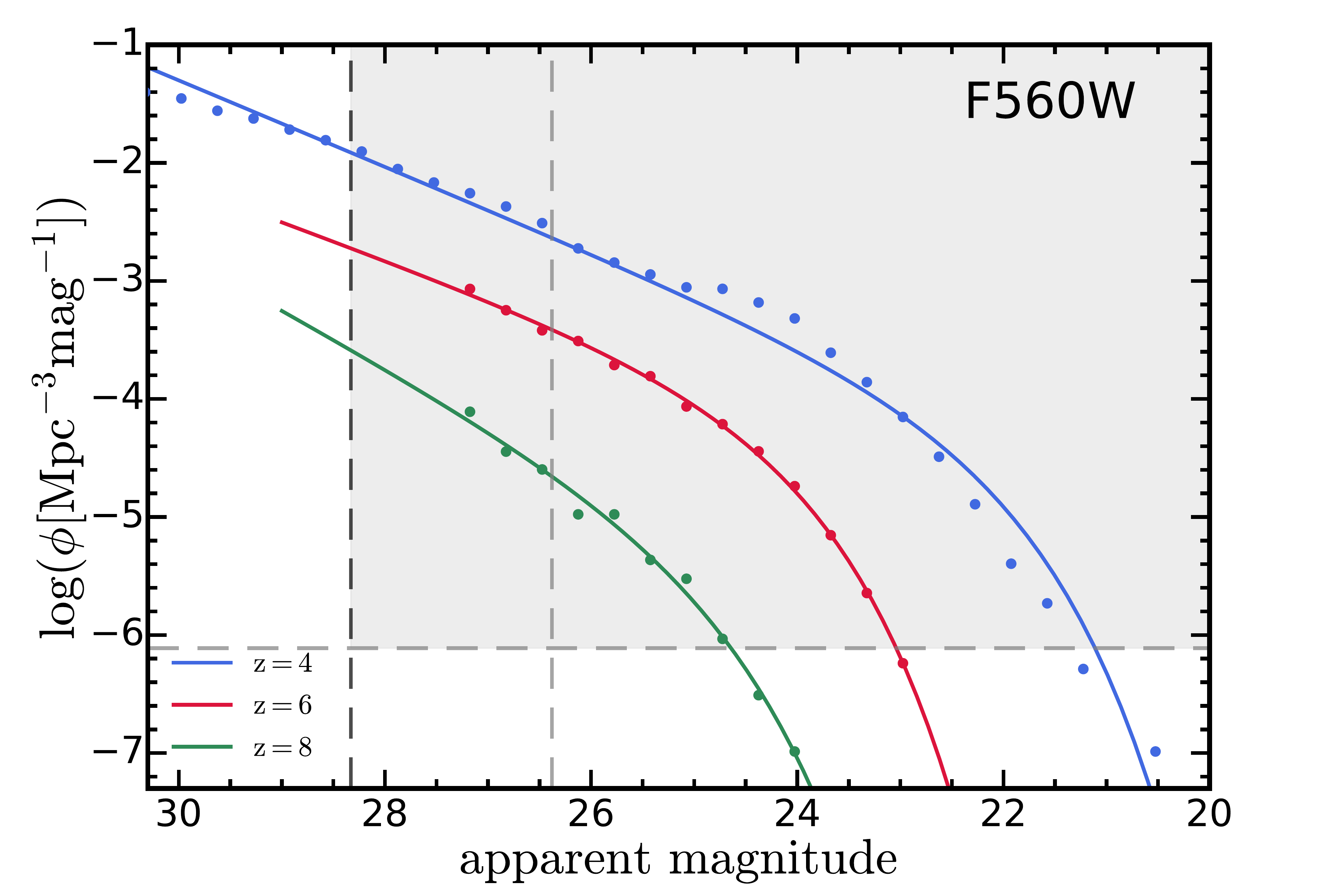}
    \includegraphics[width=0.49\textwidth]{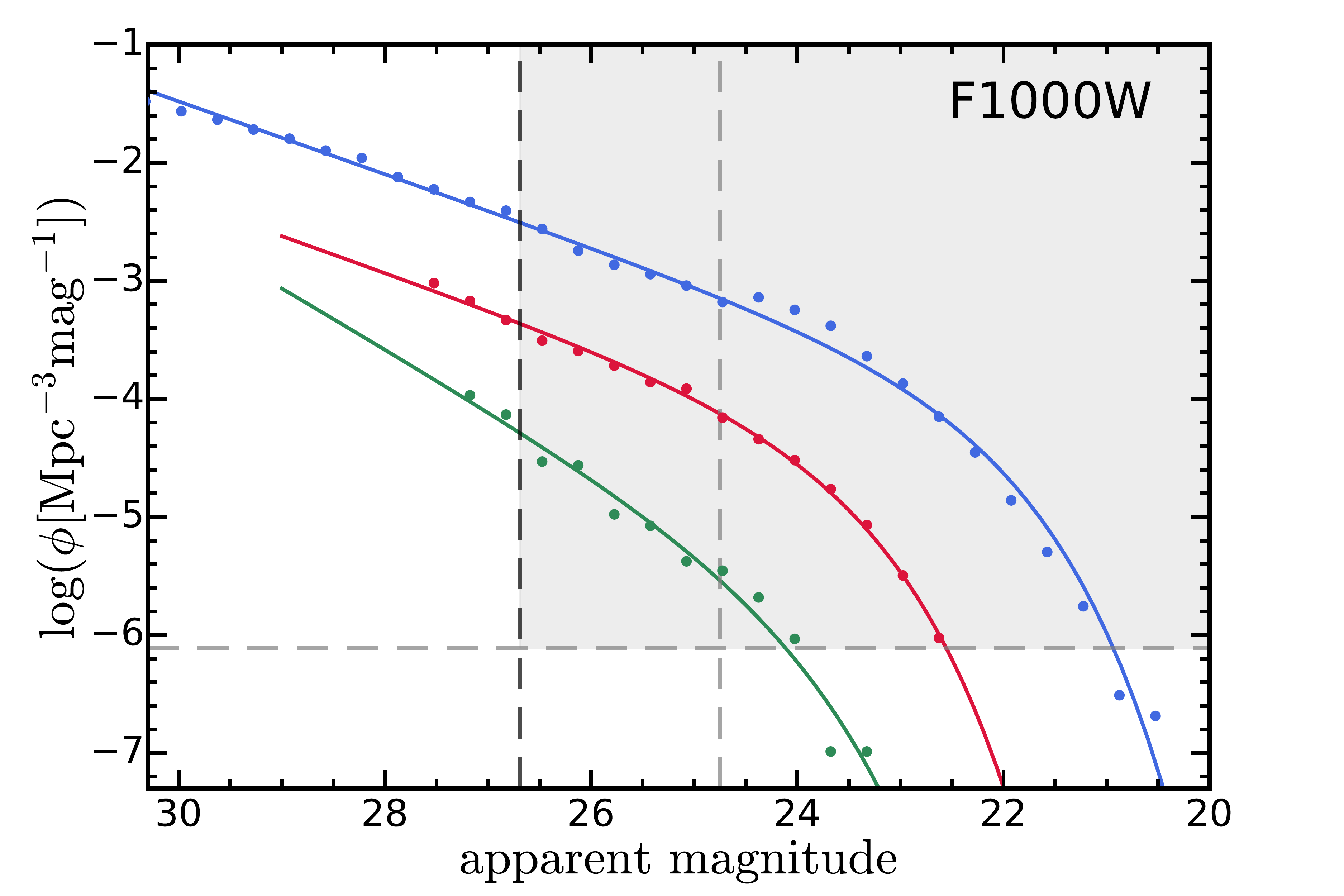}
    \includegraphics[width=0.49\textwidth]{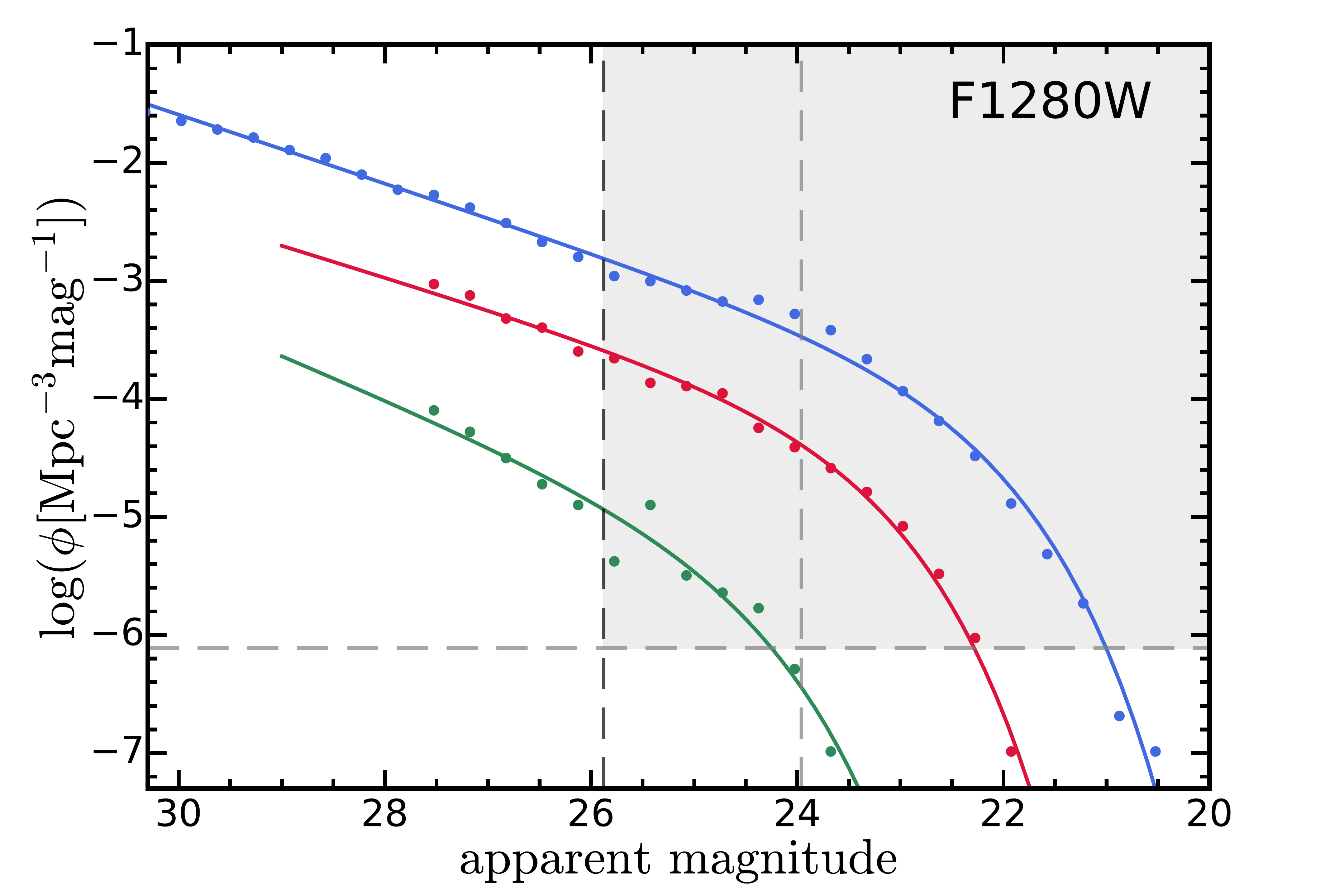}
    \includegraphics[width=0.49\textwidth]{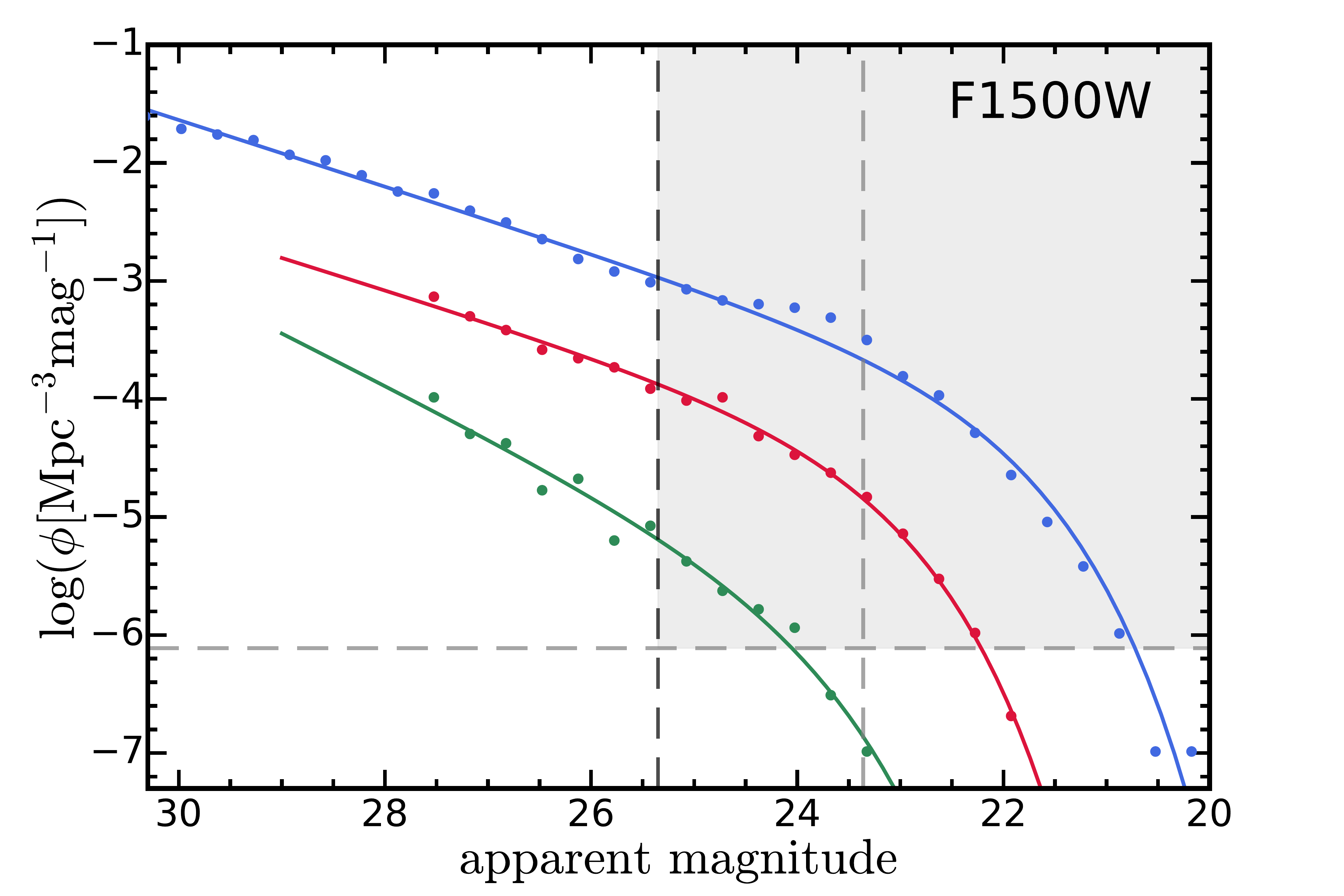}
    \caption{ \textbf{ {\it JWST} MIRI band apparent luminosity functions.} Binned estimations from simulations are shown in solid circles and the best-fit Schechter functions are shown in solid lines. The horizontal dashed line indicates the number density corresponding to one galaxy in a survey area of $500\,{\rm arcmin}^2$ with survey depth $\Delta z=1$ centered around $z=6$. Here we have assumed that all the galaxies above the detection limit can be detected and selected with $100\%$ completeness. This reference line shifts only slightly changing the survey center to $z=4$ or $z=8$. The vertical dashed lines indicate the detection limits we calculated assuming SNR$=10$, $T_{\rm exp}=10^4\,{\rm s}$ for the gray line and SNR$=5$, $T_{\rm exp}=10^5\,{\rm s}$ for the black line. }
    \label{fig:jwst_lf}
\end{figure*}

\begin{figure}
    \centering
    \includegraphics[width=0.49\textwidth]{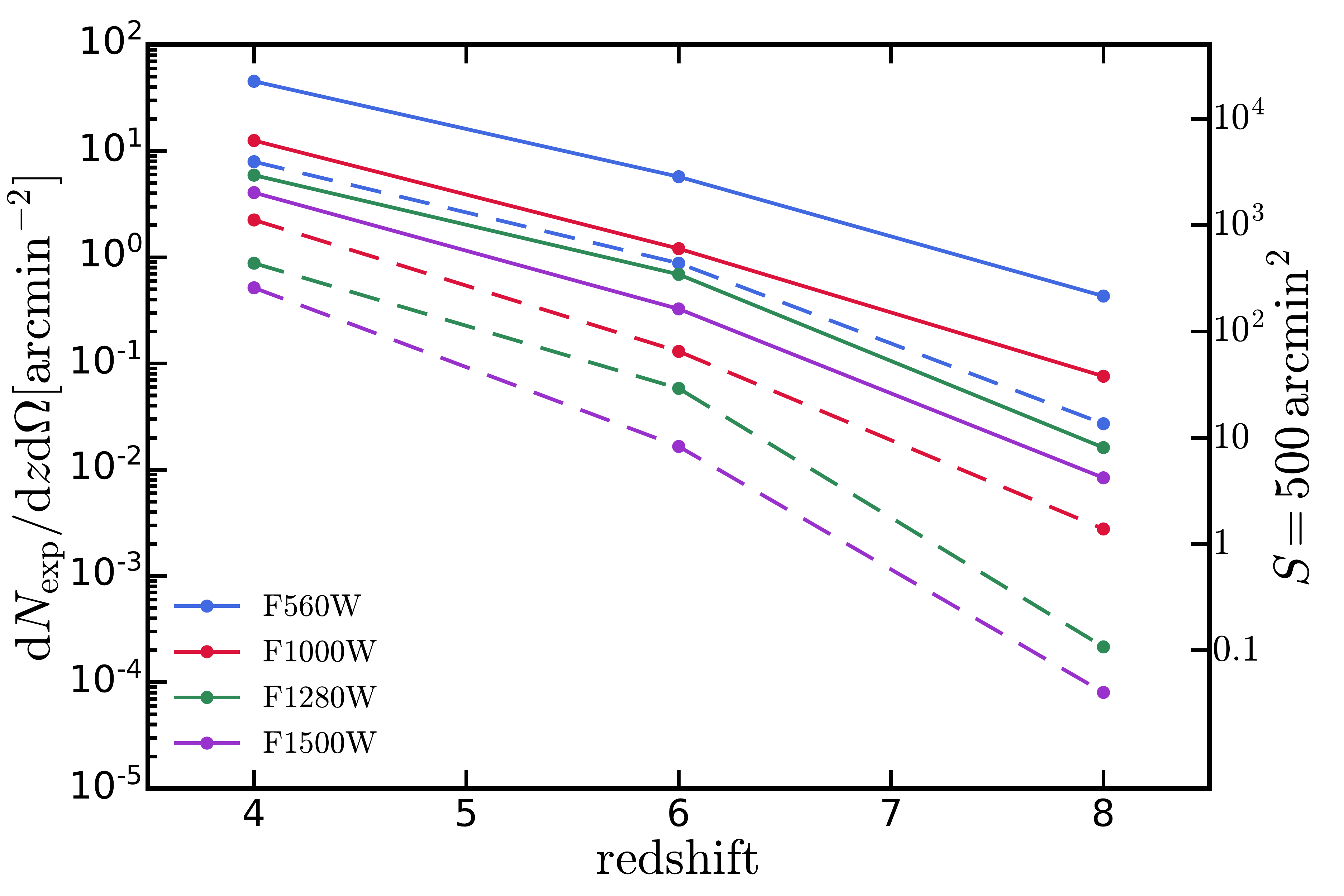}
    \caption{\textbf{Expected number counts of galaxies (theoretically estimated) in {\it JWST} MIRI bands.} We show the number of galaxies per unit area of field of view and per unit survey depth in the MIRI bands as a function of redshift. The numbers on the right hand side assume a survey area of $500\,{\rm arcmin}^2$. The dashed lines show predictions for the relatively shallow detection limit (assuming SNR$=10$, $T_{\rm exp}=10^4\,{\rm s}$) and the solid lines show predictions for the deep detection limit (assuming SNR$=5$, $T_{\rm exp}=10^5\,{\rm s}$). We refer to the results as ``theoretically estimated'' since we have assumed effectively $100\%$ completeness in detection and selection of galaxies. }
    \label{fig:jwst_nexp}
\end{figure}

The {\it JWST} MIRI provides imaging and spectroscopic observing modes from $4.9$ to $28.8\micron$ \citep{Wright2015,Rieke2015}, which could be used to identify high-redshift galaxies. In this section, we will make predictions for the apparent band luminosity functions of MIRI and numbers of galaxies expected in a survey volume.
 
\begin{table}
\caption{{\bf {\it \textbf{JWST}} MIRI wide filter characteristics and detection limits.} The table contains the pivot wavelengths, bandwidths and detection limits of the {\it JWST} MIRI wide filters involved in this work. The detection limits are calculated assuming $10^{4}{\rm s}$ and $10^{5}{\rm s}$ exposure time with target signal-to-noise ratios of $10$ and $5$ based on the {\it JWST} Exposure Time Calculator. For a given exposure time, the detection limits with target signal-to-noise ratios of $5$ to $10$ can simply be obtained through a constant shift of $\sim 0.77\mmag$.}
\centering
\begin{tabular}{ p{0.045\textwidth}|p{0.055\textwidth}|p{0.055\textwidth}|p{0.095\textwidth}|p{0.098\textwidth}}
\hline 
{\bf Filter} & wavelength & bandwidth & $m^{\rm lim}_{{\rm SNR=10}, T_{\rm exp}=10^4{\rm s}}$ & $m^{\rm lim}_{{\rm SNR=5}, T_{\rm exp}=10^5{\rm s}}$  \\
 & [$\mu{\rm m}$] & [$\mu{\rm m}$] & [${\rm mag}$] & [${\rm mag}$]\\
\hline
\hline
{\bf F560W}  & 5.6  & 1.2 & 26.38 & 28.33 \\
{\bf F770W}  & 7.7  & 2.2 & 25.68 & 27.61 \\
{\bf F1000W} & 10.0 & 2.0 & 24.75 & 26.69 \\
{\bf F1130W} & 11.3 & 0.7 & 23.74 & 25.68 \\
{\bf F1280W} & 12.8 & 2.4 & 23.96 & 25.88 \\
{\bf F1500W} & 15.0 & 3.0 & 23.36 & 25.35$^{\rm a}$ \\
{\bf F1800W} & 18.0 & 3.0 & 22.33$^{\rm a}$ & 24.12$^{\rm a}$\\
{\bf F2100W} & 21.0 & 5.0 & <22$^{\rm b}$   &  <24$^{\rm b}$ \\
{\bf F2500W} & 25.0 & 4.0 & <22$^{\rm b}$   &  <24$^{\rm b}$ \\
\hline
\end{tabular}
\raggedright \\
$^{\rm a}$ The number of exposures per specification is doubled while the number of groups per integration is halved, to avoid saturation of background exposure (see text for details). \\
$^{\rm b}$ The saturation of background exposure cannot be avoided by tuning the observational strategy, so rough upper limits are given.
\label{tab:jwst_filters}
\end{table}

Basic information of the MIRI broad bands are listed in Table~\ref{tab:jwst_filters} along with the detection limit of each band. The detection limits are calculated using the {\it JWST} Exposure Time Calculator (ETC)\footnote{\url{https://jwst.etc.stsci.edu/}} with the following configuration details. Sources are treated as point sources, and the exposure time is set to either $10^{4}\rm s$ or $10^{5}\rm s$ as indicated in the table. Furthermore, the readout pattern is set to SLOW, which yields a high signal-to-noise ratio and can efficiently reach a  maximum survey depth. For the $10^{4}{\rm s}$ exposure time we employ the full sub-array with $18$ groups per integration, $1$ integration per exposure, $24$ exposures per specification. For the $10^{5}{\rm s}$ exposure time we employ $42$ groups per integration, $1$ integration per exposure, $100$ exposures per specification. For some long wavelength bands, to avoid saturation of background exposure, we double the number of exposures per specification while halving the groups per integration. The aperture used for imaging is circular with radius $0.1"$. The background subtraction is performed with a sky annulus with inner and outer radii of $1.2"$ and $1.98"$. The ETC background model includes celestial sources (zodiacal light, interstellar medium, and cosmic IR background) along with telescope thermal and scattered light. This background model varies with the target coordinates (RA, Dec) and time of year. Here, we choose position on the sky to the Hubble Ultra-Deep Field at RA = 03:32:39.00, Dec = -27:47:29.0 and choose the background configuration to be ``low''. For each band, we then set up all the ETC parameters as described and then vary the apparent magnitude of the source until a target signal-to-noise ratio is reached. This then sets the corresponding apparent magnitude detection limit for this band for various exposure times and target signal-to-noise ratios. 

From the simulation, we derive the apparent band luminosity of galaxies following the procedure in \citetalias{Vogelsberger2020}. The luminosity functions in each band from different simulations are combined and the resulting combined luminosity functions can be well reproduced by the Schechter function~\citep[][]{Schechter1976}
\begin{equation}
\phi(M)= \dfrac{{\rm d}n}{{\rm d}M} = =\dfrac{0.4\ln{(10)}\,\phi^{\ast}}{10^{0.4(M-M^{\ast})(\alpha+1)}}e^{-10^{-0.4(M-M^{\ast})}},
\label{eq:schechter}
\end{equation}
where $M$ is the rest-frame (apparent) magnitude when describing rest-frame (apparent) luminosity functions. The Schechter function can also be expressed as
\begin{equation}
\phi(L)= \dfrac{{\rm d}n}{{\rm d}\log{L}} =\phi^{\ast} \Big(\dfrac{L}{L_{\ast}}\Big)^{\alpha+1} e^{-L/L_{\ast}},
\end{equation}
when describing the number density of galaxies per dex of luminosity, where $\phi^{\ast}$ is the number density normalization, $L_{\ast}$ is the break luminosity and $\alpha$ is the faint-end slope. The best-fit Schechter function parameters are shown in Appendix~\ref{appsec:schfit}. In Figure~\ref{fig:jwst_lf}, we compare the binned estimations and Schechter fits of the apparent band luminosity functions of four selected {\it JWST} MIRI broadbands to the detection limits of MIRI in Table~\ref{tab:jwst_filters}. This comparison demonstrates the promise of MIRI for the identification of galaxies up to $z\sim 8$, assuming a survey area~\footnote{For reference, the survey area of the Spitzer Extended Deep Survey \citep[SEDS,][]{Ashby2013} in the IRAC $3.6$ and $4.5\micron$ bands is about $1.46\, {\rm deg}^2$, which was followed by the Spitzer-Cosmic Assembly Deep Near-infrared Extragalactic Legacy Survey \citep[S-CANDELS,][]{Ashby2015} with a smaller area of $0.16\, {\rm deg}^2$ ($576\,{\rm arcmin}^2$) and increased depth.} of $\sim 500\,{\rm arcmin}^2$ and depth of $\Delta z=1$. 

Based on the Schechter fit of the luminosity function, the cumulative number density of galaxies can be calculated by integrating the Schechter function as
\begin{align}
\label{eq:phi_cum}
\phi_{\rm cum}\left (<M^{\rm lim}\right )&=\int^{\infty}_{L^{\rm lim}}\!\!\!\!\phi^{\ast}\left(\dfrac{L}{L^{\ast}}\right)^{\alpha}\exp\left(-\dfrac{L}{L^{\ast}}\right)\dfrac{{\rm d}L}{L^{\ast}} \nonumber \\
&=\phi^{\ast}\,\Gamma_{\rm inc}(\alpha+1,10^{-0.4(M^{\rm lim}-M^{\ast})}),
\end{align}
where $\Gamma_{\rm inc}(a,z)=\int^{\infty}_{z}t^{a-1}e^{-t}{\rm d}t$ is the incomplete upper Gamma function and $M^{\rm lim}$ is the magnitude limit of integration. Then the expected number of galaxies above the detection limit ($m^{\rm lim}$) in a survey volume can be calculated as
\begin{equation}
N_{\rm exp}  \simeq \phi_{\rm cum}(<m^{\rm lim}) \,\, \dfrac{{\rm d}V_{\rm com}}{{\rm d}\Omega \, {\rm d}z}(z) \,\, \Delta \Omega \,\, \Delta z,
\end{equation}
where $\Delta \Omega$ is the solid angle corresponding to the area of the survey, $\Delta z$ is the redshift coverage of the survey and ${\rm d}V_{\rm com}/{\rm d}\Omega{\rm d}z$ is the differential comoving volume element at the redshift of the survey
\begin{equation}
\dfrac{{\rm d}V_{\rm com}}{{\rm d}\Omega\,{\rm d}z}(z)=\dfrac{c\,(1+z)^{2}\,\,d_{\rm A}(z)^{2}}{H_{0}\,E(z)},
\end{equation}
where $d_{\rm A}(z)$ is the angular diameter distance and $H(z) = H_{0}E(z)$ is the Hubble parameter at redshift $z$. We note that the calculations here effectively assume that all the galaxies above the magnitude limit $m_{\rm lim}$ can be detected and selected in a real imaging survey. However, in reality, the completeness correction is necessary to recover the physical abundance of galaxies, the expected number of galaxies detected should be
\begin{equation}
N_{\rm exp} \simeq \left( \int_{-\infty}^{+\infty} {\rm d}m\, \dfrac{{\rm d}\phi_{\rm cum}(<m)}{{\rm d}m} \, P(m,z)\right) \,\dfrac{{\rm d}V_{\rm com}}{{\rm d}\Omega \, {\rm d}z}(z) \,\, \Delta \Omega \,\, \Delta z,
\end{equation}
where $P(m,z)$ is the completeness function for a specific observation. Our calculations above effectively assume $P(m,z)=0$ when $m>m_{\rm lim}$ and $P(m,z)=1$ when $m<=m_{\rm lim}$. But in a real observation, the transition of $P(m,z)$ at $m_{\rm lim}$ would be smooth. The shape of the completeness function depends on the details of the observational configuration, background noises and selection criteria. Therefore, the expected numbers calculated here should be interpreted as a theoretical estimate and one needs to be cautious comparing them with real observational results.

\begin{figure}
    \centering
    \includegraphics[width=0.49\textwidth]{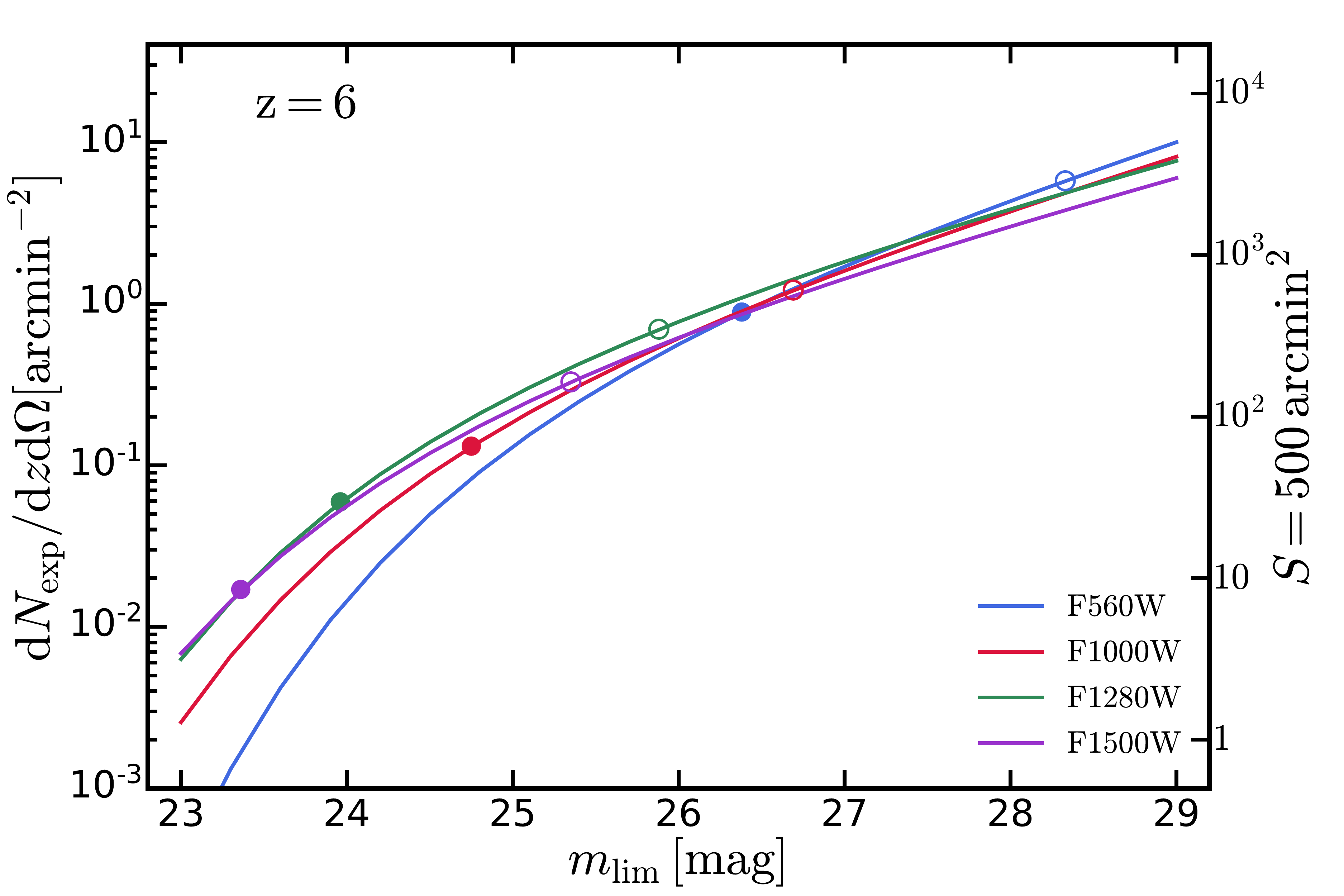}
    \caption{\textbf{Expected number counts of galaxies (theoretically estimated) in {\it JWST} MIRI bands versus detection limit.} We show the number of galaxies per unit area of field of view and per unit survey depth in the MIRI bands as a function of detection limit at $z=6$. The detection limit is expressed as the limiting apparent magnitude. The labelling is the same as Figure~\ref{fig:jwst_nexp}. The solid (open) circles indicate the number counts at the detection limit assuming SNR$=10$ and $T_{\rm exp}=10^4\,{\rm s}$ (SNR$=5$ and $T_{\rm exp}=10^5\,{\rm s}$). We note that the detection limit does not scale with the total exposure time in a trivial way and also depends on the details of the observational configuration.
    }
    \label{fig:jwst_nexp_vs_mlim}
\end{figure}

The expected number counts of galaxies (theoretically estimated) in the four selected MIRI broadbands at $z=4\operatorname{-}8$ are shown in Figure~\ref{fig:jwst_nexp}. At $z=6$, assuming a target SNR$=5$ and $T_{\rm exp}=10^5\,{\rm s}$, $\sim 3000$ ($\sim 500$) galaxies are expected in F560W (F1000W) with a survey area of $\sim 500\,{\rm arcmin}^2$ and depth of $\Delta z=1$. Even at $z=8$ in F1500W (which has the poorest sensitivity among the four bands selected), $\sim 5$ galaxies are expected to be detected with the same observational configuration. Previously, the deepest large NIR galaxy surveys~\citep[e.g.,][]{Ashby2013,Steinhardt2014,Ashby2015} have been mainly conducted by the Infrared Array Camera \citep[IRAC;][]{Fazio2004} on the Spitzer Space Telescope \citep{Werner2004}. In these surveys, the $5\operatorname{-}\sigma$ survey depth is around $25 \mmag$ in the IRAC $3.6$ and $4.5\micron$ bands~\citep{Steinhardt2014,Mortlock2017,Stefanon2017} and around $24.5 \mmag$ in the IRAC $5.8$ and $8.0\micron$ bands~\citep[deepest data in the GOODS-N field as discussed in][]{Stefanon2017}. Compared to the detection limits of {\it JWST} MIRI with comparable exposure time ($10^{5}\,{\rm s} \sim 30\,{\rm hr}$), the detection limit at $5-10\micron$ can be pushed deeper by {\it JWST} by about $2\mmag$ and the number of galaxies selected at $z \gtrsim 6$ with complete photometric data at rest-frame UV to NIR $8\micron$ will be boosted from of order ten~\citep{Stefanon2017} to a few hundred. In Figure~\ref{fig:jwst_nexp_vs_mlim}, we show the expected number counts of galaxies at $z=6$ detected in the four MIRI broadbands as a function of detection limit. Relevant numbers can be read out when the detection limits vary with other observational configurations. For shallow detection limits, the number counts are lower in MIRI bands located at shorter wavelengths due to stronger dust attenuation at shorter wavelengths in rest-frame NIR. The difference diminishes for deep detection limits since faint, low-mass galaxies dominate the total number counts and they are less affected by dust attenuation.

\subsection{Bolometric IR luminosity functions and obscured SFRD}

The FIR luminosities of galaxies are dominated by dust continuum emission, which is reprocessed from the UV emission of young stellar populations. Therefore, FIR luminosities are sensitive to on-going star formation in galaxies. Since the IR SEDs of galaxies peak at FIR wavelengths, the bolometric IR luminosity integrated at $8\operatorname{-}1000~\micron$ has often been used as an indicator for star formation~\citep[e.g.,][]{Kennicutt1998,Murphy2011}. In particular, IR based measurements can reveal the star formation in heavily dust obscured galaxies which may be too faint to be observed in the rest-frame UV and optical. The IR measurements could reflect a significant portion of the total amount of star formation in galaxies. Therefore, the constraints of the bolometric IR luminosity functions as well as the abundance of the most luminous IR emitters (usually accompanied by strong dust obscuration) are important to constrain the cosmic SFRD.

\begin{figure}
    \centering
    \includegraphics[width=0.49\textwidth]{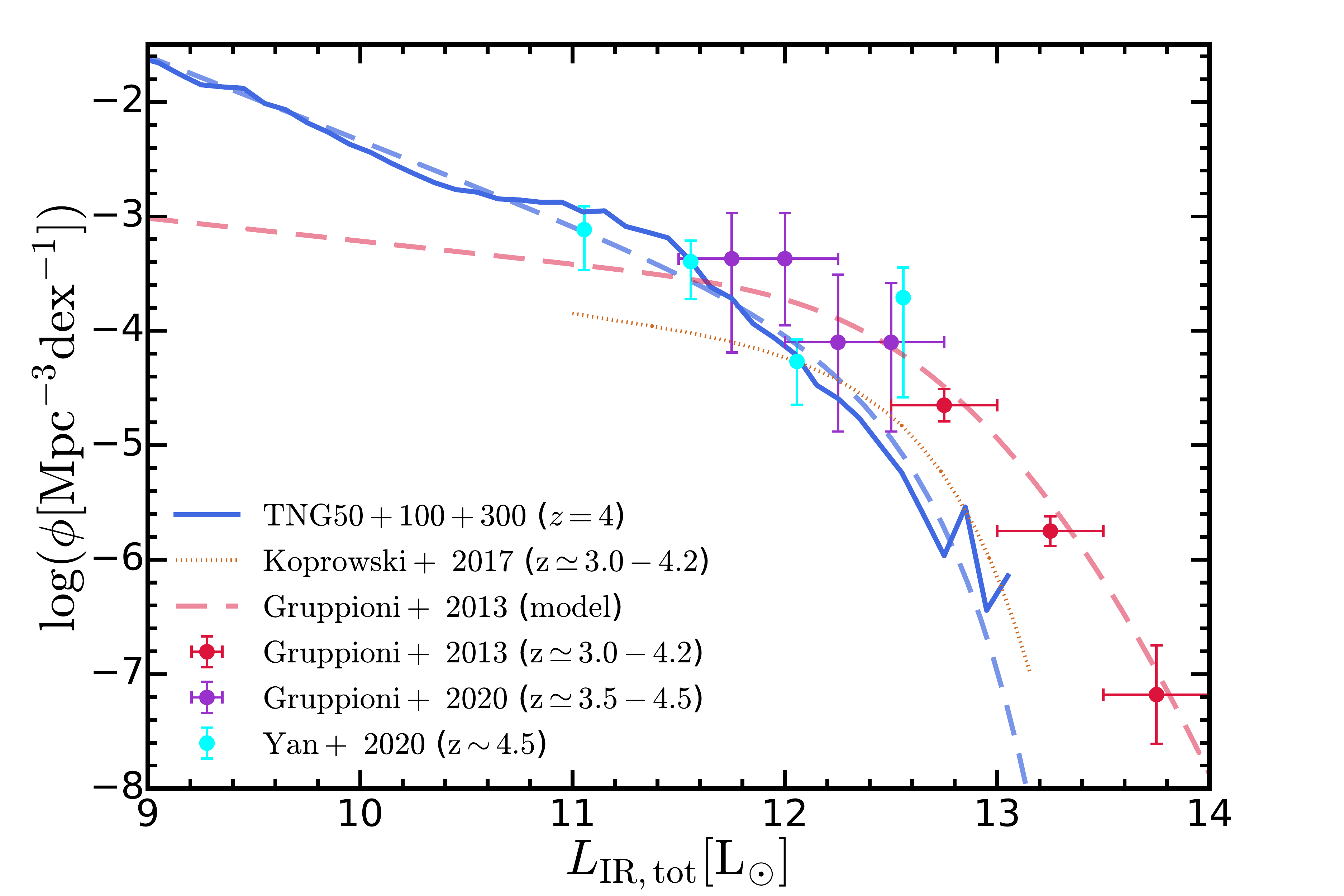}
    \includegraphics[width=0.49\textwidth]{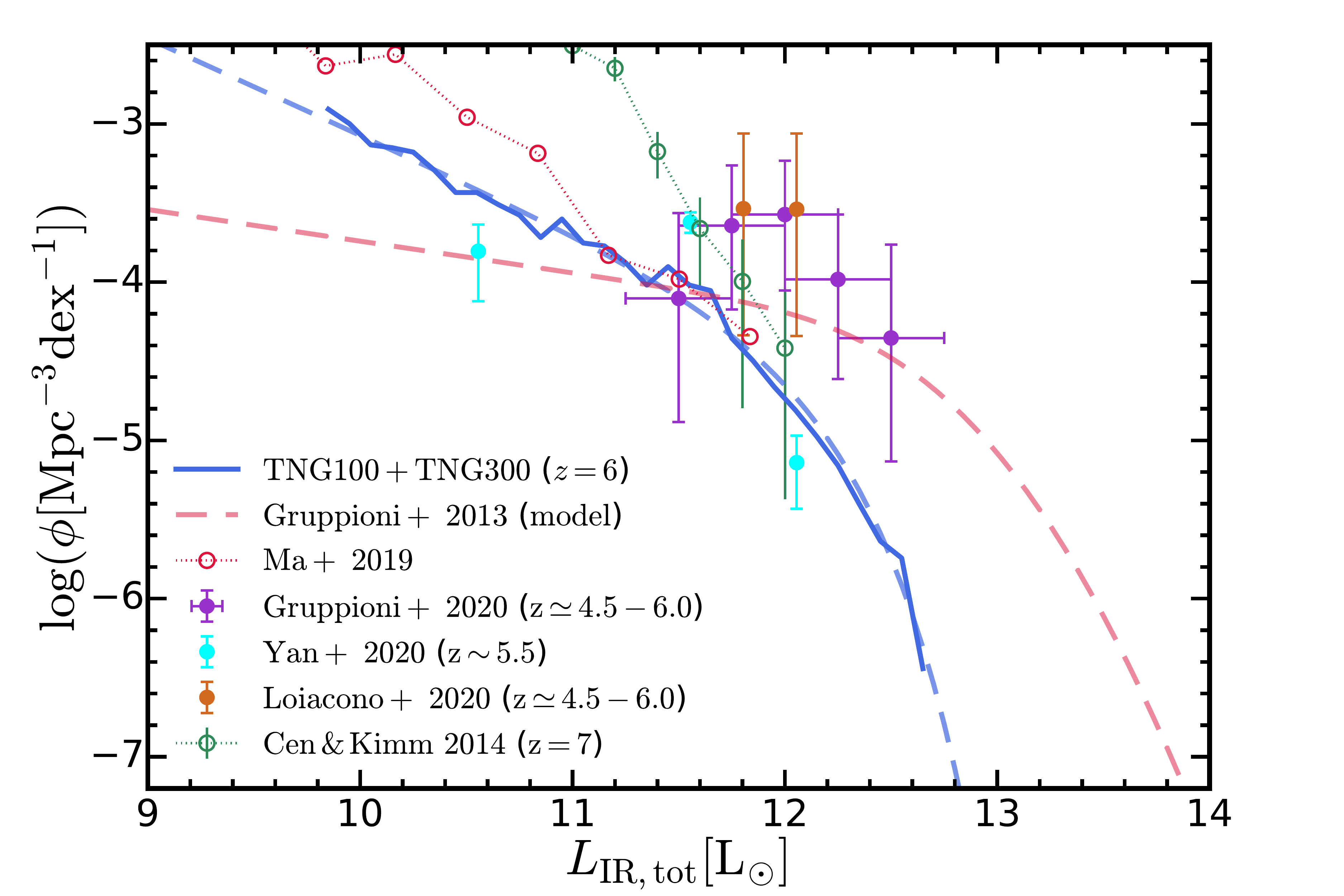}
    \caption{\textbf{Bolometric IR luminosity function.} The galaxy bolometric IR luminosity functions at $z=4$ and $z=6$ from the IllustrisTNG simulations are presented in solid lines. Binned estimations and Schechter fits from observations~\citep{Gruppioni2013,Koprowski2017,Gruppioni2020} are shown with solid markers and dashed lines. Predictions from theoretical works~\citep{Cen2014,Ma2019} are shown with open markers and dotted lines. Compared to observations, TNG underpredicts the abundance of most luminous IR galaxies at both redshifts and predicts a steeper faint-end slope.  }
    \label{fig:bol-lf}
\end{figure}

\begin{figure*}
    \centering
    \includegraphics[width=0.49\textwidth]{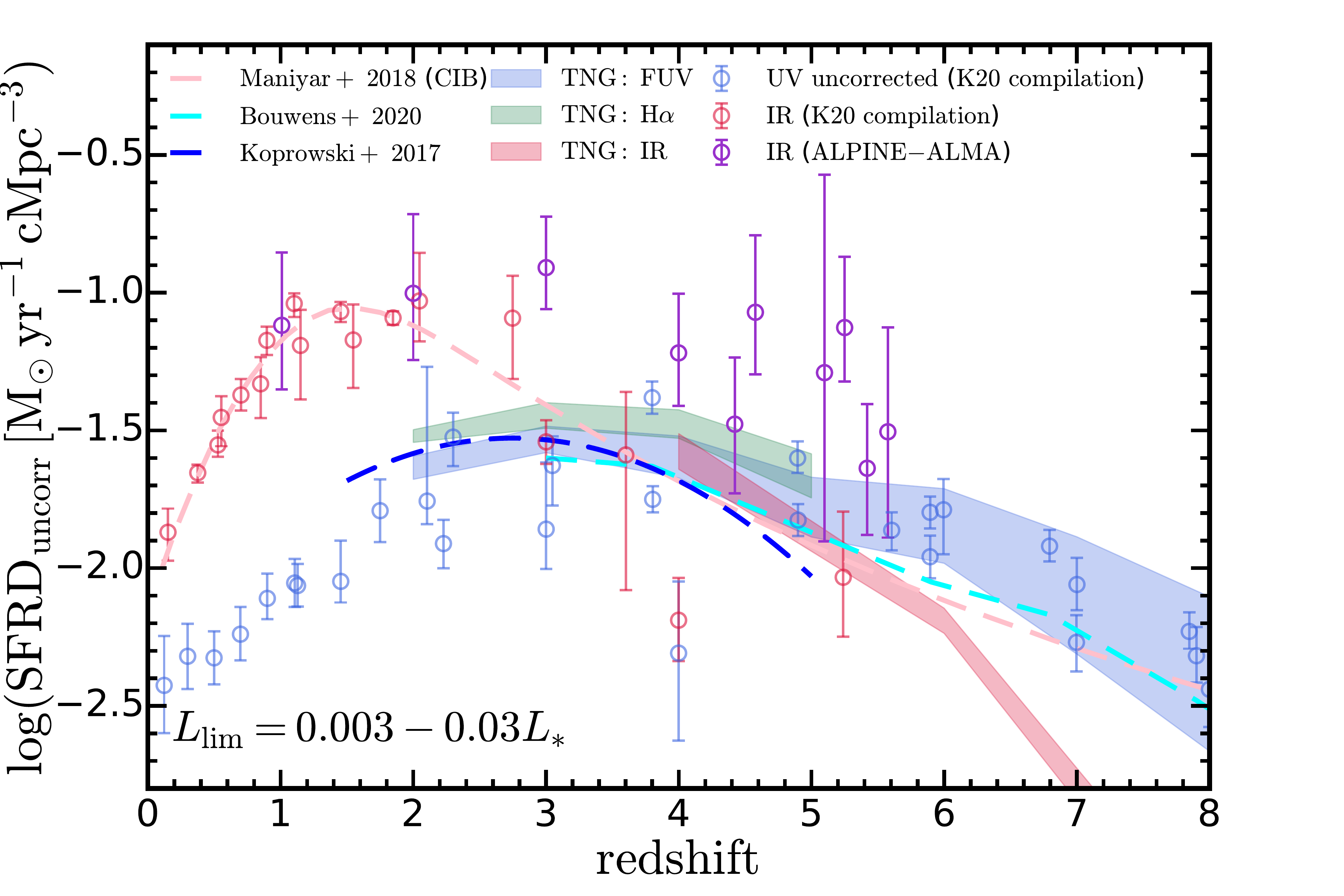}
    \includegraphics[width=0.49\textwidth]{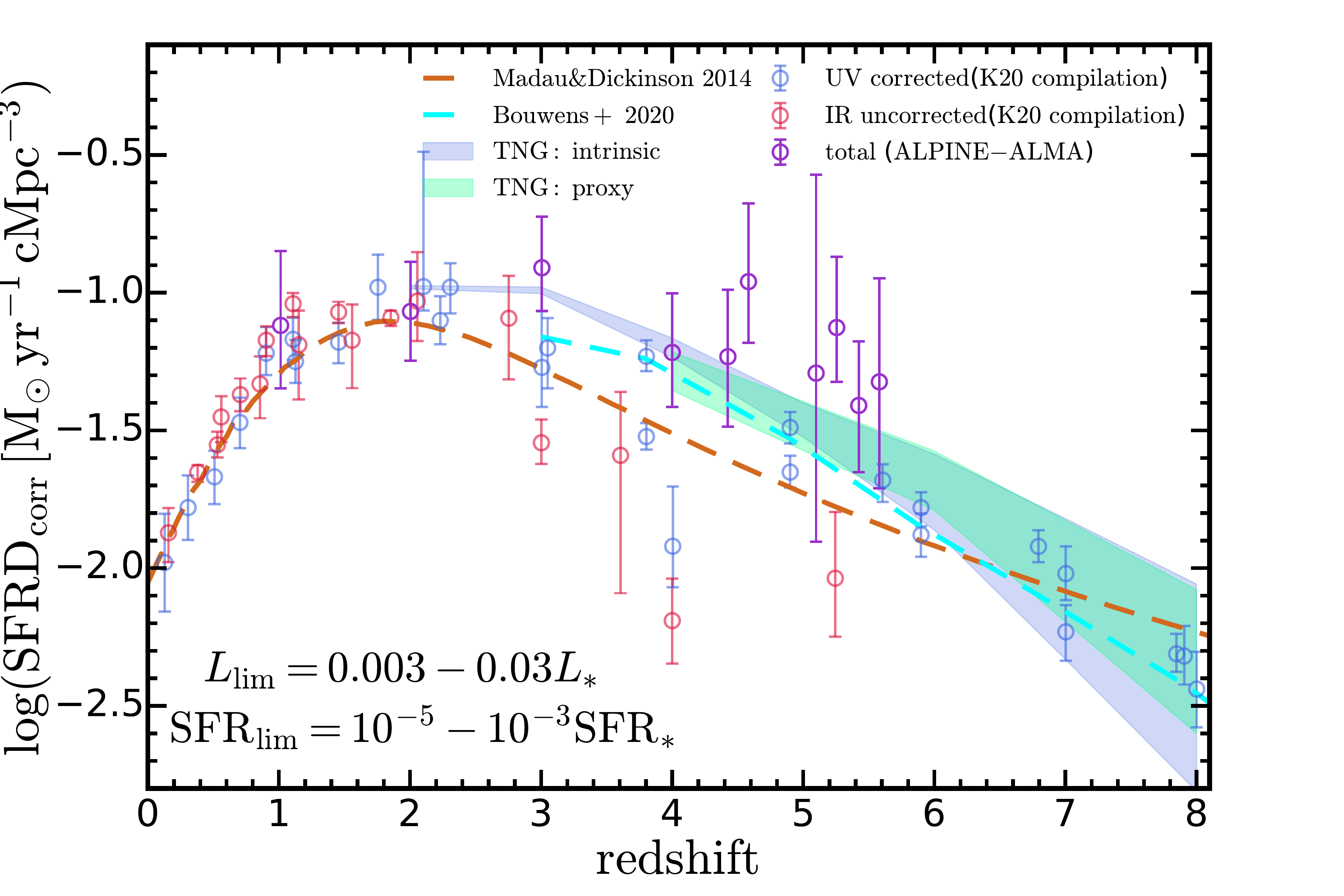}
    \caption{\textbf{Evolution of the cosmic SFRD.} In the left panel, we show the uncorrected SFRD estimated from the UV, ${\rm H}_{\alpha}$ and IR luminosity densities of galaxies in the IllustrisTNG simulations. The UV and ${\rm H}_{\alpha}$ luminosities are based on results in \citetalias{Vogelsberger2020} and \citetalias{Shen2020}. The luminosity densities are derived by integrating the luminosity functions to $0.003\operatorname{-}0.03\,L_{\ast}$, where $L_{\ast}$ is the break luminosity. The luminosity densities are converted to SFRD assuming the conversion factors in \citet{Murphy2011}. We compare the simulation results with observations compiled in \citet{Khusanova2020}. The UV (IR) observations are shown with blue (red) points. Specifically, we show the most recent IR-based estimations from \citet{Khusanova2020,Loiacono2020,Gruppioni2020} in purple points. For reference, we show the evolution of SFRD constrained by UV observations from \citet{Koprowski2017,Bouwens2020} in dashed lines. In the right panel, we show the total SFRD (corrected for dust attenuation). The light and dark blue dashed lines show constraints from \citet{Bouwens2020} and \citet{Koprowski2017} based on compiled UV observations. In the right panel, we show the total SFRD as a function of redshift. From simulations, we calculate it through two approaches: the summation of SFRDs inferred by UV and IR indicators (labelled with ``proxy''), and the SFRD measured from the instataneous SFR of gas cells in simulations (labelled with ``intrinsic''). For observations, we show the SFRD measurements corrected for dust attenuation. The predicted SFRD from simulations is lower than the results of the ALPINE-ALMA survey, in particular the obscured SFRD.
    }
    \label{fig:sfrd}
\end{figure*}

\subsubsection{Bolometric IR luminosity function}
\label{sec:results-irlf}

In Figure~\ref{fig:bol-lf}, we present the galaxy bolometric IR luminosity function at $z=4$ and $z=6$ from the IllustrisTNG simulations. The luminosity functions from different simulations are combined together and resolution corrections are not applied since we find that the TNG50, TNG100 and TNG300 results are consistent with each other in shared dynamical ranges. We compare the simulation predictions with the following observations: the Herschel luminosity functions~\citep[e.g.,][]{Gruppioni2013}; the SCUBA-2 luminosity functions~\citep[e.g.,][]{Koprowski2017}; the ALPINE-ALMA luminosity functions~\citep[e.g.,][]{Gruppioni2020}. We also include the bolometric IR luminosity functions converted from the [C\Rmnum{2}] luminosity functions of UV-selected galaxies in \citet{Yan2020,Loiacono2020} by \citet{Gruppioni2020}. At $z\simeq4$, the Herschel and the ALPINE-ALMA luminosity function measurements are mutually consistent with each other and cover complementary luminosity ranges. However, the SCUBA-2 luminosity function is systematically lower, potentially due to the different method in calculating the IR luminosities and incompleteness in sample selection. The TNG prediction is in agreement with the Herschel and ALPINE-ALMA luminosity functions at $L_{\rm IR} \lesssim 10^{12}\Lsun$ while underpredicting the abundance of galaxies at the bright end with $L_{\rm IR} \gtrsim 10^{13}\Lsun$. The actual number of IR luminous galaxies is important for the cumulative IR luminosity density of galaxies, since the faint-end slopes of IR luminosity functions are usually shallow ($\alpha \gg -2$). In addition, at the faint end, TNG predicts steeper luminosity functions than the extrapolation of observational based models. If we translate the typical IR luminosity ($L_{\rm IR} \sim 10^{13}\Lsun $) of these ``missing'' galaxies in TNG simulations to SFR, it would roughly be $\sim 1000\msun/{\rm yr}$ ~\citep{Kennicutt1998,Murphy2011}. Assuming the averaged properties of high-redshift ($z\gtrsim 3$) sub-millimeter galaxies with a typical specific star-formation rate of $\log{\rm sSFR} \sim -8.5$~\citep{Magnelli2012,daCunha2015,Miettinen2017}, the implied stellar masses of these ``missing'' galaixes will be $M_{\ast} \sim 10^{11.5} \msun$. 

At $z\simeq6$, the TNG prediction is consistent with the observations at $L_{\rm IR}\lesssim 10^{11.5}\Lsun$ while it is lower at $L_{\rm IR} \gtrsim 10^{12}\Lsun$ by about an order of magnitude. The difference of the redshift bins in observations and simulations will likely not lead to such a large discrepancy, since the evolution of the number density normalization and the break luminosity is slow at $z\gtrsim 2$~\citep{Gruppioni2020}. At the bright end, similar to the discrepancy at $z=4$, the simulation underpredicts the abundance of galaxies at $L_{\rm IR} \gtrsim 10^{12}\Lsun$ and predicts a steeper faint-end slope compared to the evolutionary model constrained in \citet{Gruppioni2013}. We also compare our results with theoretical predictions from \citet{Cen2014} (radiative transfer calculations on $198$ galaxies with $5\times 10^{8}\msun< M_{\ast} < 3\times 10^{10}\msun$ at $z=7$) and \citet{Ma2019} (radiative transfer on FIRE-2 high-redshift simulations using {\sc Skirt}). Their predictions are in good agreement with TNG at $L_{\rm IR} \sim 10^{12}\Lsun$ while suggesting much steeper faint-end luminosity functions, which deviate from observations further. At the bright end, the simulations by \citet{Cen2014} and \citet{Ma2019} lose predictive power since the most massive, heavily obscured galaxies were not sampled. If we extrapolate their IR luminosity functions to brighter luminosities (by either a power-law or exponential cut-off), both of them will infer a lower abundance of luminous IR galaxies compared to observations, similar to the TNG prediction. 

In observational studies, merger-driven starbursts have been considered as the classical explanation for the extremely high IR luminosities of sub-millimeter galaxies \citep[e.g.,][]{Chakrabarti2008,Engel2010,Narayanan2010,Hayward2011}. Since our galaxy selection and radiative transfer calculations are all performed based on subhaloes identified by the {\sc Subfind} algorithm, it is possible that we mistreat merging systems (with high intrinsic SFRs and IR luminosities) as distinct, individual merging galaxies and therefore underpredict the abundance of IR-luminous systems. To test this scenario, we first select all the galaxies with ${\rm SFR}\geq 100\msun/{\rm yr}$ (which roughly corresponds to $L_{\rm IR}\gtrsim 10^{12}\Lsun$) and $M_{\ast}\geq 100 \times m_{\rm b}$ at $z=4$ in TNG300. Based on this sample of galaxies, we start from the most massive galaxy (as the host galaxy) and link galaxy companions with distance to host $d\leq d_{\rm lim} = 50\pkpc$ to this host galaxy. We repeat the same process to the remaining set of galaxies that haven't been linked to other galaxies. Finally, after all the galaxies are properly linked to their hosts, the SFRs of galaxies are recalculated by summing the SFRs of all the galaxy companions linked to the host galaxies. We find that the typical enhancement in SFR of the host galaxies is $\Delta ( \log{\rm SFR} ) = 0.2^{+0.08}_{-0.10}$ while the abundance of galaxies with ${\rm SFR}\geq 1000\msun/{\rm yr}$ (which roughly corresponds to $L_{\rm IR}\gtrsim 10^{13}\Lsun$) does not change at all. Even if we increase $d_{\rm lim}$ to $1\pMpc$, the abundance of galaxies with ${\rm SFR}\geq 1000\msun/{\rm yr}$ only increases by about $0.2\,{\rm dex}$ which is still far from explaining the underprediction we found in the bolometric IR luminosity function. So we conclude that the underprediction is unlikely related to the definition and selection of galaxies in post-processing.

This underpredicted abundance of luminous IR galaxies in TNG is consistent with the underpredicted UV dust attenuation in massive galaxies and the underpredicted abundance of heavily-obscured UV red systems found in \citetalias{Shen2020}. \citet{Hayward2020} has also reported the underpredicted counts of sub-millimeter galaxies in TNG. Since our model is calibrated based on the dust-attenuated UV luminosity functions, the solution to the tension would not only require a higher abundance of dust (stronger dust attenuation) but also higher intrinsic UV emission (either stronger star formation or additional radiation sources). One plausible explanation is resolution effects. Though resolution corrections have been considered in our model, the predictions at the bright (massive) end are still primarily determined by TNG100 and TNG300, which may fall short of resolving star formation and metal enrichment in the dense ISM. A similar effect has been investigated in \citet{Lim2020}, where the star formation efficiency of proto-clusters in TNG300 is much lower than observations and part of it can be attributed to resolution effects. However, the luminosity functions (of all the bands we studied) of TNG100 and TNG300 at $z\geq 4$ are roughly identical at the bright (massive) end~\citep{Vogelsberger2020}, indicating that convergence (of galaxy bulk luminosities and masses) is reached in massive galaxies at the resolution level in the redshift range. Another possible explanation is the sub-grid stellar/AGN feedback model adopted in TNG. In \citet{Hayward2020}, it is shown that the original Illustris simulation with an older feedback model predicts the correct abundance of sub-millimeter galaxies, and massive galaxies in TNG may be quenched too early by feedback. Though it is hard to isolate the exact cause, the abundance of bright IR galaxies remains an appealing channel to constrain the feedback model in cosmological simulations.

\subsubsection{Obscured star formation at high redshift}

To relate the luminosity functions to the obscured SFRD in the Universe, we first perform fits to the bolometric IR luminosity functions with the Schechter function (Equation~\ref{eq:schechter}, the best-fit parameters are shown in Appendix~\ref{appsec:schfit}) and integrate best-fit Schechter functions to derive the cumulative IR luminosity density
\begin{align}
\label{eq:L_cum}
\rho_{\rm IR}\left (<L_{\rm IR}^{\rm lim}\right ) & = \int^{\infty}_{L_{\rm IR}^{\rm lim}}\!\!\!\!\phi^{\ast} L_{\ast} \left(\dfrac{L_{\rm IR}}{L^{\ast}}\right)^{\alpha+1}\exp\left(-\dfrac{L_{\rm IR}}{L^{\ast}}\right)\dfrac{{\rm d}L_{\rm IR}}{L^{\ast}} \nonumber \\
&=\phi^{\ast}\,L_{\ast}\,\Gamma_{\rm inc}(\alpha+2,L_{\rm IR}^{\rm lim}/L_{\ast}),
\end{align}
where $\Gamma_{\rm inc}(a,z)$ is the incomplete upper Gamma function as in Equation~\ref{eq:phi_cum} and $L_{\rm IR}^{\rm lim}$ is the limit of integration which will be discussed later. The derived cumulative IR luminosity density is converted to the ${\rm SFRD}_{\rm IR}$ (the SFRD measured in IR) using the calibration in \citet{Murphy2011}, which assumed the \citet{Chabrier2003} initial mass function consistent with the choice in TNG. We apply the same procedure to the predicted UV luminosity functions in \citetalias{Vogelsberger2020} and the ${\rm H}\alpha$ luminosity functions in \citetalias{Shen2020} to derive ${\rm SFRD}_{\rm UV}$ and ${\rm SFRD}_{{\rm H}\alpha}$. We finally derive the SFRD traced by the three indicators and we sum up ${\rm SFRD}_{\rm IR}$ and ${\rm SFRD}_{\rm UV}$ to get the total SFRD inferred from indicators~\footnote{In principle, both UV and ${\rm H}\alpha$ light trace the unobscured star formation in galaxies. The dust attenuation of them could be different due to the geometry of radiation source distribution. Since the majority of the energy absorbed by dust is in UV, it is better to pair ${\rm SFRD}_{\rm UV}$, rather than ${\rm SFRD}_{{\rm H}\alpha}$, with ${\rm SFRD}_{\rm IR}$ to calculate total SFRD.}.

In the left panel of Figure~\ref{fig:sfrd}, we compare these predictions with the observational constraints compiled in \citet[][see references therein]{Khusanova2020}. These observational results have been converted to the \citet{Chabrier2003} initial mass function. These observations have been divided into three groups: blue circles, the SFRD derived from FUV observations without dust attenuation corrections; red circles, SFRD derived from IR and sub-millimeter observations; purple circles, ${\rm SFRD}_{\rm IR}$ derived from the recent ALPINE-ALMA survey~\citep{Khusanova2020,Loiacono2020,Gruppioni2020}. The predictions from TNG\footnote{We note that the continuous evolution of the IR luminosity density as well as the corresponding SFRD is obtained through interpolation of the simulation and post-processsing results at $z=4,6,8$.} are presented as shaded regions with the lower (upper) boundary corresponding to $0.03$ ($0.003$) $L_{\ast}$ as the integration limits, where $L_{\ast}$ is the break luminosity of the best-fit Schechter function. These integration limits are commonly adopted in literature. The ${\rm SFRD}_{\rm UV}$ predicted from TNG is in agreement with the majority of the UV observations at $z=2\operatorname{-}8$. The lower boundary of the prediction agrees almost perfectly with the most recent constraint from \citet{Bouwens2020}, where $0.03\,L_{\ast}$ was adopted as the integration limit. The uncertainty induced by the integration limit is small compared to the observational uncertainties at $z\lesssim 5$, but becomes significant at $z\gtrsim 6$ due to the steep faint-end slope of the luminosity function there. In addition, the ${\rm SFRD}_{{\rm H}\alpha}$ predicted from TNG agrees with the ${\rm SFRD}_{\rm UV}$ with $\lesssim 0.2\, \mathrm{dex}$ differences. Such differences could be attributed to the uncertainties in modelling the emission line intensity of young stellar populations as discussed in \citetalias{Shen2020}. The ${\rm SFRD}_{\rm IR}$ predicted from TNG agrees with the model in \citet{Maniyar2018}, derived based on the cosmic infrared background (CIB) anisotropies, at $z\lesssim 6$ but is lower at $z\gtrsim 6$. Compared to the recent ALPINE-ALMA observations, the TNG prediction is about half an order of magnitude lower. The deficiency is related to the underpredicted abundance of luminous IR galaxies in TNG, compared to the IR luminosity functions constrained by the ALPINE-ALMA survey, as shown above. The prediction from TNG supports the picture that the obscured SFRD dominates at low redshift, starts to decline at $z\sim 2$ and eventually becomes subdominant compared to the unobscured SFRD at $z\gtrsim 5$. 

\begin{figure}
    \centering
    \includegraphics[width=0.49\textwidth]{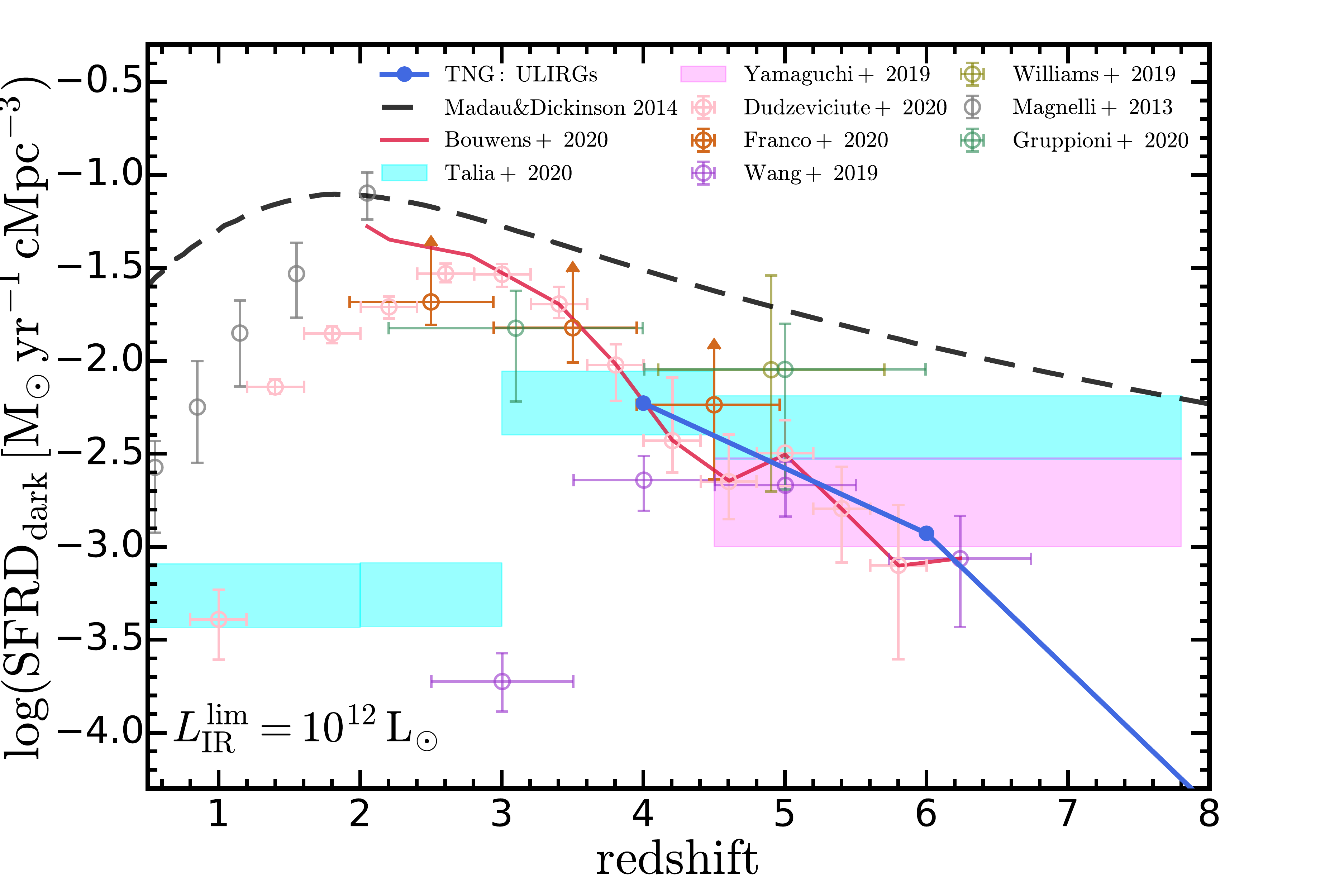}
    \caption{\textbf{Cosmic SFRD contributed by the optical-NIR-dark galaxies.} We show the SFRD contributed by galaxies with $L_{\rm IR} > 10^{12}\Lsun$ (ULIRGs) in TNG through integration over the predicted bolometric IR luminosity function. This serves as an upper limit for the SFRD of optical-NIR-dark galaxies. We compare the results with the observational constraints compiled in \citet{Bouwens2020} as well as the constraints from the ALPINE-ALMA survey \citep{Gruppioni2020,Talia2020}, as labelled. The TNG prediction is in agreement with some of the observations at $z\geq 4$ while being $\sim 1\sigma$ lower than the estimations from \citet{Williams2019,Gruppioni2020,Talia2020}.}
    \label{fig:sfrd_ulirg}
\end{figure}

In the right panel of Figure~\ref{fig:sfrd}, we show the total SFRD predicted from the IllustrisTNG simulations and compare it with the SFRD measured in the IR and UV that have been corrected for dust attenuation. To calculate the total SFRD from simulations, the first way is to simply sum up ${\rm SFRD}_{\rm UV}$ and ${\rm SFRD}_{\rm IR}$ derived via cumulative luminosity densities. An alternative way is to measure the SFRD from the instantaneous star formation in gas cells in simulations. We determine the SFR of each TNG galaxy by summing up the SFRs in gas cells within a $30\pkpc$ aperture of the galaxy. The choice of aperture is consistent with the box size we set for radiative transfer post-processing and our definition for other galaxy bulk properties. We perform resolution corrections to the SFRs of galaxies in TNG100 and TNG300 using the method described in \citetalias{Vogelsberger2020} and \citetalias{Shen2020}. With the resolution corrected SFRs of galaxies, we construct SFR functions and combine the SFR functions of TNG50, TNG100 and TNG300 as in \citetalias{Vogelsberger2020} and \citetalias{Shen2020}. Finally, we fit the combined SFR functions with the Schechter function and integrate the Schechter function to the integration limit, $10^{-3}$ ($10^{-5}$) ${\rm SFR}_{\ast}$, where ${\rm SFR}_{\ast}$ is the break SFR of the SFR function. The limit is smaller compared to the integration limits we choose when integrating over the luminosity function, because the break SFR of the SFR functions is relatively larger and empirically we find these limits make the SFRD consistent with observations simultaneously at $z=2$ and $z=8$. As shown in the right panel of Figure~\ref{fig:sfrd}, the total SFRDs derived by the two methods agree remarkably well and give similar level of uncertainties induced by integration limits. Compared to the observational constraints, the total SFRD from TNG is larger than the \citet{Madau2014} model at $2\lesssim z\lesssim6$ while it gives similar prediction at $z\gtrsim 7$. The lower boundary of the prediction agrees well with the recent estimation in \citet{Bouwens2020} (assuming $L_{\rm lim} = 0.03\,L_{\ast}$) where an updated dust correction was applied and the contribution of heavily obscured, optical/NIR dark galaxies was accounted for. However, the predicted SFRD is still $\sim 0.2 \, \mathrm{dex}$ lower than what is suggested by the ALPINE-ALMA survey.

To illustrate the contribution of heavily obscured, luminous IR galaxies to the cosmic SFRD, we show the TNG prediction and several direct/indirect observational constraints of the SFRD in optical/NIR dark galaxies in Figure~\ref{fig:sfrd_ulirg}. We show the fiducial model adopted in \citet{Bouwens2020} and the compiled observational data therein, including the SFRD of ULIRGs in \citet{Magnelli2013} (integrated to $L_{\rm IR}=10^{12}\Lsun$) and \citet{Dud2020} (integrated to $S_{\rm 870} = 1{\rm mJy}$), constraints of NIR dark galaxies in \citet{Yamaguchi2019, Williams2019, Wang2019, Franco2020}. In addition, we include constraints from the ALPINE-ALMA survey~\citep{Talia2020, Gruppioni2020}. The TNG prediction is derived by integrating the bolometric IR luminosity function to $L_{\rm IR}=10^{12}\Lsun$, which roughly corresponds to ${\rm SFR}>100\msun/{\rm yr}$. The criterion matches the property of typical DSFGs that are missed from optical and NIR observations~\citep[e.g.,][]{Yamaguchi2019,Wang2019,Talia2020}. Since this essentially assumes that all the galaxies with $L_{\rm IR}=10^{12}\Lsun$ are optical/NIR dark, the prediction should be taken as an upper limit. Among observations, despite the consensus on the steady decline of the SFRD contributed by luminous IR galaxies at $z\gtrsim 2$, the quantitative contribution varies in the literature, with roughly an order of magnitude variation at $z\gtrsim 4$. The TNG prediction is in agreement with some of the observations at $z\geq 4$ while in tension with the results from \citet{Williams2019,Gruppioni2020,Talia2020} at a $1\sigma$ level. Considering that the TNG prediction is an upper limit for the SFRD of optical/NIR dark galaxies, the discrepancy is even larger than revealed by this comparison. The discrepancy is related to the underprediction of luminous IR galaxies in TNG as discussed above and demonstrates that the difference in the total SFRD between TNG predictions and observations is mainly due to a deficiency of luminous IR galaxies in TNG.

\subsection{Dust temperature}

Most of the IR and sub-millimeter emission from galaxies is produced through thermal emission of dust grains in the ISM, heated by the UV radiation of young stellar populations (if not accounting for AGN activity and CMB heating). The volume averaged thermal emission of dust is often described by a modified blackbody (MBB) spectrum, $f_{\nu} \sim  \epsilon_{\nu} B_{\nu}(T_{\rm dust})$, where $B_{\nu}(T_{\rm dust})$ is the blackbody spectrum and $\epsilon_{\nu} \sim \nu^{\beta}$ is the volume averaged emissivity function that accounts for the dust opacity (in the optical thin limit). At FIR wavelengths, the emissivity index $\beta$ takes the value about $2$ according to dust scattering theory~\citep{Draine2001} and consistent with observational constraints~\citep[e.g.,][]{Dunne2000,Draine2007}. $T_{\rm dust}$ represents the temperature of the relatively cool dust component which dominates the emission at long wavelengths (as opposed to the warm dust that dominates mid-IR emission). The peak of a MBB spectrum ($L_{\nu}$) is related to the so-called ``peak'' dust temperature as
\begin{equation}
    \lambda_{\rm peak} = 96.64 \left(\frac{T_{\rm peak}}{30{\rm K}}\right)^{-1} \micron\, ,
    \label{eq:lpeak-T}
\end{equation}
assuming $\beta = 2$. This manifests as the Wien's displacement law, but here the peak is defined as where $L_{\nu}$ reaches a maximum. This is a direct and less model-dependent way to measure dust temperature based solely on the observed FIR spectrum of galaxies and is widely used in observational studies~\citep[e.g.,][]{daCunha2013,Schreiber2018,Faisst2020}. To determine the peak of the FIR spectrum ($L_{\nu}$), we first find where the maximum flux is reached in the {\sc Skirt} output spectrum. Since the wavelength grid for {\sc Skirt} calculations is sparse, to determine the peak more accurately, we interpolate the galaxy spectrum with a cubic spline and determine the peak based on the interpolated function. %We refer to the temperature derived in this way as $T_{\rm peak}$. 

A more robust way in determining the dust temperature is through SED modelling. Considering the warm dust emission in the mid-IR and dust opacity (not in the optically thin limit), the dust continuum emission can be described as~\citep{Blain2003,Casey2012b}
\begin{align}
    & L_{\nu}(\lambda) = N_{\rm bb} f(\lambda,\beta,T_{\rm sed}) + N_{\rm pl} (\lambda/\lambda_{\rm c})^{\alpha} e^{-(\lambda/\lambda_{\rm c})^{2}}, \nonumber \\
    & N_{\rm pl} = N_{\rm bb} f(\lambda_{\rm c},\beta,T_{\rm sed}), \nonumber \\
    & f(\lambda,\beta,T_{\rm sed}) = \dfrac{ \big(1-e^{-(\lambda_{\rm 0}/\lambda)^{\beta}} \big) (\lambda/c)^3 }{e^{hc/\lambda k T_{\rm sed}} - 1},
    \label{eq:lir_model}
\end{align}
where $\alpha$ is the mid-IR spectral slope, $\lambda_{\rm c}$ is related to $\alpha$~\citep[as defined in][and taking a typical value of $\sim 50\micron$]{Casey2012b}, $N_{\rm bb}$ is a normalization factor, $\lambda_{\rm 0}$ is the wavelength where the optical depth is unity, $T_{\rm sed}$ is the SED temperature. The free parameters of the model are $\alpha$, $\beta$, $T_{\rm sed}$ and $N_{\rm bb}$. These parameters can be determined by fitting the IR spectrum from {\sc Skirt} calculations with the model. $T_{\rm sed}$ is usually larger than $T_{\rm peak}$ especially when the optically-thin assumption is no longer valid~\citep{Casey2012b}. We note that both $T_{\rm sed}$ and $T_{\rm peak}$ should be considered as light-weighted dust temperatures, as opposed to the mass-weighted temperature of cold dust emitting at the Rayleigh-Jeans tail of the FIR spectrum, and are more appropriate for comparing to observational constraints. The CMB is an additional source of dust heating, especially at high redshift when the CMB temperature is close to the dust temperature, which is not captured by the simulations. To account for that, a CMB correction of dust temperatures is implemented as \citep[e.g.,][]{daCunha2013}
\begin{equation}
    T_{\rm dust} = \Big( T_{\rm dust, i}^{4+\beta} + T_{\rm CMB}^{4+\beta}(z) - T_{\rm CMB}^{4+\beta}(0) \Big)^{1/(4+\beta)}.
\end{equation}
where $T_{\rm dust, i}$ is the intrinsic dust temperature and $T_{\rm CMB}(z) = T_{\rm CMB}(0)\,(1+z)=2.73\,{\rm K}\,\,(1+z)$ is the CMB temperature at $z$. In addition, since the dust continuum of high-redshift galaxies is always measured against the CMB background, the detectability as well as the measured FIR fluxes are affected by CMB. However, this effect has already been corrected in most of the observational studies.

\begin{figure}
    \centering
    \includegraphics[width=0.49\textwidth]{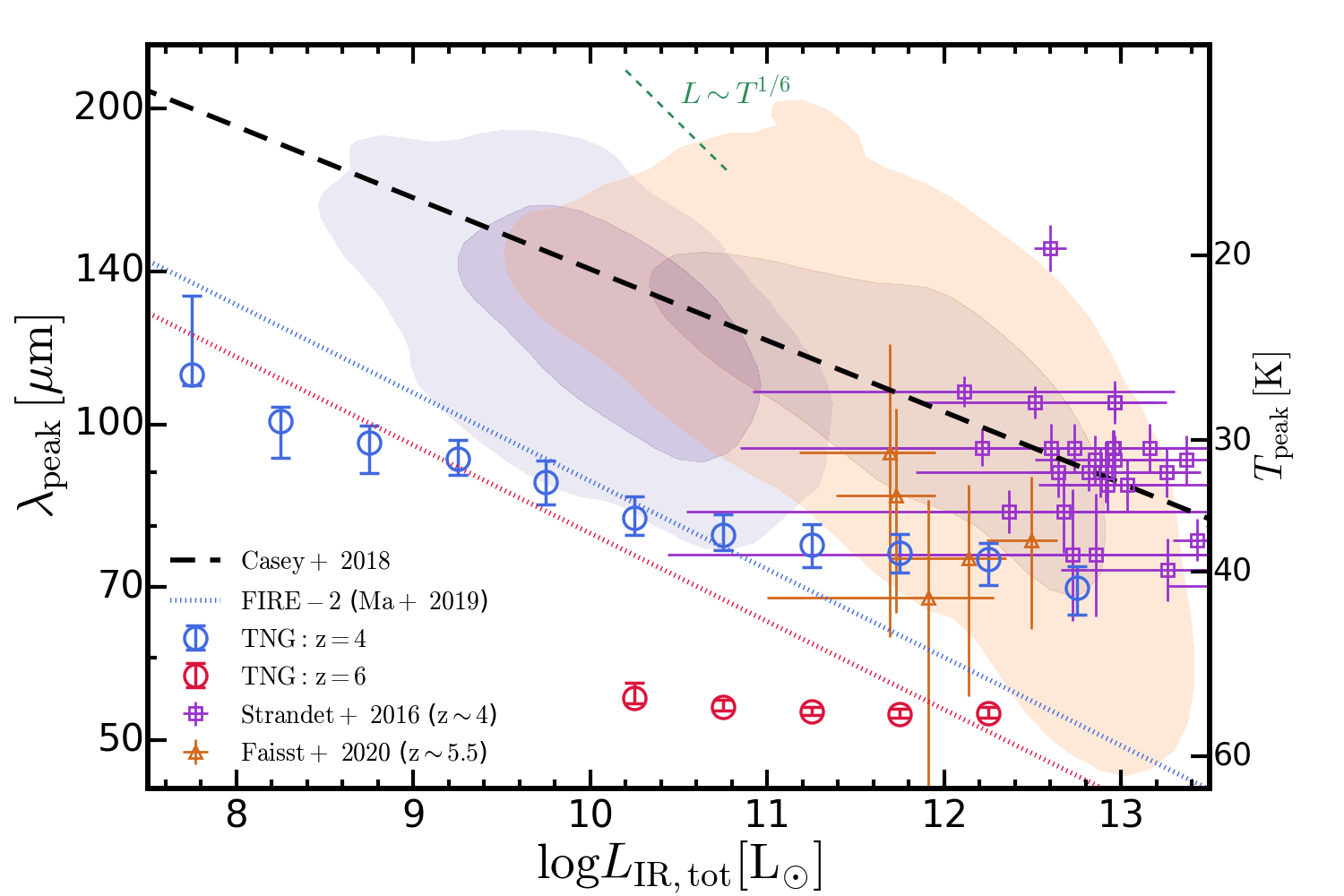}
    \caption{ \textbf{Peak of IR spectrum (dust peak temperature) versus bolometric IR luminosity.} We show the peak of the IR spectrum versus bolometric IR luminosity from TNG in open circles, with error bars indicating $1.5\sigma$ dispersion ($87\%$ inclusion) in each bin. The peak of the spectrum is related to the peak dust temperature $T_{\rm peak}$ with Equation~\ref{eq:lpeak-T}. We compare the results with observational constraints based on: H-ATLAS samples at $z<0.1$ in \citet{Valiante2016} (purple shaded region, $1$ and $2\sigma$ contour); COSMOS samples at $0.5<z<2$ from \citet{Lee2013} (orange shaded region); SPT-detected DSFGs at $z\sim 4$ in \citet{Strandet2016} (purple squares); ALMA observations of four main-sequence galaxies at $z\sim 5.5$ from \citet{Faisst2020} (orange triangles). The former three are compiled by \citet{Casey2018} and a suggested redshift-independent relation from this work is shown in the black dashed line. For reference, theoretical predictions from the FIRE-2 simulations~\citep{Ma2019} are shown in dotted lines. The TNG predicted IR peak wavelengths of $z\geq 4$ galaxies are shorter than the low-redshift galaxies and thus disfavor a redshift-independent $\lambda_{\rm peak}$-$L_{\rm IR}$ relation. The peak wavelength anti-correlates with IR luminosity until a plateau of dust temperature is reached at $L_{\rm IR}\gtrsim 10^{11}\Lsun$. TNG predictions at $z=4$ are compatible with the SPT-selected galaxies at $z \sim 4$ while the predictions at $z=6$ give higher peak dust temperatures than suggested by $z\sim 5.5$ galaxies in \citet{Faisst2020}.}
    \label{fig:tdust_vs_lir}
\end{figure}

\subsubsection{$T_{\rm dust}$ versus $L_{\rm IR}$}

In Figure~\ref{fig:tdust_vs_lir}, we show the peak of the IR spectrum (and $T_{\rm peak}$ equivalently) as a function of bolometric IR luminosity of TNG galaxies. The peak of the spectrum is translated to the peak dust temperature $T_{\rm peak}$ using Equation~\ref{eq:lpeak-T}. Only galaxies with $M_{\ast}>1000m_{\rm b}$ ($m_{\rm b}$ is the baryon mass resolution) are considered in this analysis to reduce the random scatter caused by poor sampling of radiation sources. We compare the results with observational constraints computed by \citet{Casey2018} based on the H-ATLAS samples at $z\lesssim 0.1$ from \citet{Valiante2016}, the COSMOS samples at $0.5\lesssim z\lesssim 2$ from \citet{Lee2013} and the South Pole Telescope (SPT) detected DSFGs at $z\sim 4$ from \citet{Strandet2016}. In addition, we include results from ALMA observations of four main-sequence galaxies at $z\sim 5.5$ reported in \citet{Faisst2020}. Theoretical predictions from the FIRE-2 simulations~\citep{Ma2019} are shown for reference. Our results disfavor the redshift-independent $\lambda_{\rm peak}$-$L_{\rm IR}$ relation proposed by \citet{Casey2018}. Similar to the predictions from FIRE-2, at low IR luminosities, the peak wavelengths at $z=4$ in TNG are about two times lower than the value in the local Universe. The peak wavelength anti-correlates with IR luminosity but the slope of the relation is shallower than the $\lambda_{\rm peak} \sim T_{\rm peak}^{-1} \sim L_{\rm IR}^{-1/(4+\beta)}$ law (assuming $\beta=2$) due to increasing abundance of dust in more luminous IR galaxies. When $L_{\rm IR}$ reaches $\sim 10^{11}\Lsun$, a plateau of $T_{\rm peak} \simeq 40\,{\rm K}$ is reached. This is likely due to the increasing dust opacity in more IR luminous galaxies which makes the hot dust hidden from observation and effectively decreases the light-weighted dust temperature. This plateau feature makes the TNG prediction more consistent with the SPT-detected galaxies at $z\sim 4$, compared to the FIRE-2 results. The TNG predicted peak dust temperature at $z=6$ is higher than that at $z=4$ by about $20\,{\rm K}$ and is higher than the dust temperatures of galaxies observed at $z\sim 5.5$ \citep{Faisst2020}. In addition, the scatter of the predicted dust temperature is also smaller than galaxies in observations. This is likely caused by the sparseness of the wavelength grid in {\sc Skirt} calculations, which makes the spectral peaks cluster around grid points.

\begin{figure}
    \centering
     \includegraphics[width=0.49\textwidth]{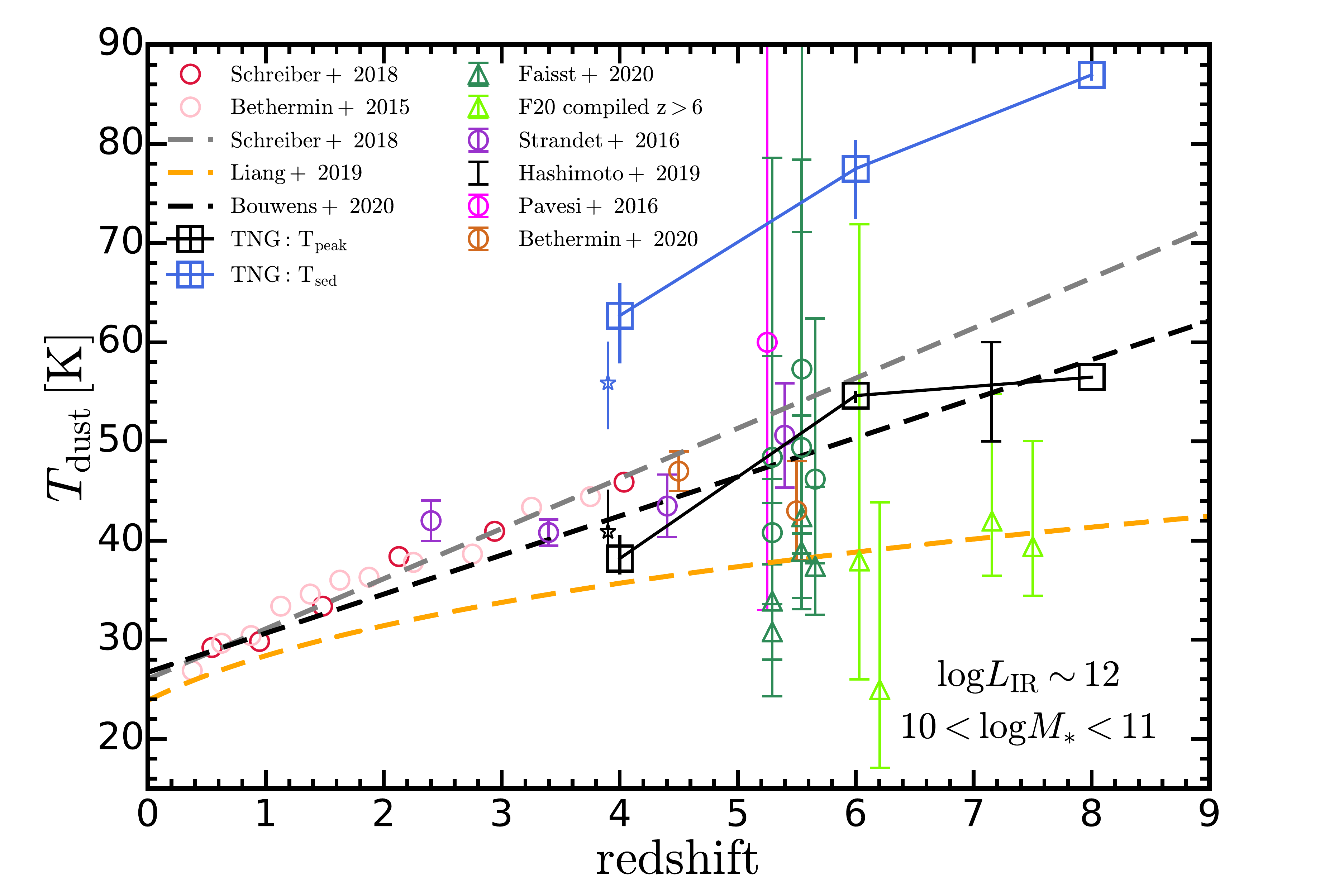}
    \caption{ \textbf{Dust temperature versus redshift.} We show the dust temperature as a function of redshift. The TNG predictions are the median $T_{\rm peak}$ and $T_{\rm sed}$ of galaxies with $10^{10}\msun < M_{\ast} < 10^{11}\msun$ in all three simulations. The error bars indicate $1.5\sigma$ scatter. With open stars, we show the median and scatter of only TNG50 galaxies at $z=4$. For comparison, we show observational constraints from \citet{Schreiber2018} (including the \citealt{Bethermin2015} samples), \citet{Strandet2016}, \citet{Faisst2020} (and $z>6$ galaxies in \citealt{Knudsen2016,Hashimoto2019,Bakx2020} re-measured). In addition, we include constraints from \citet{Pavesi2016,Hashimoto2019,Bethermin2020}. Observations classified as peak dust temperature measurements are shown in triangles. Others are shown in circles. For reference, we show relations proposed by \citet{Schreiber2018} and \citet{Bouwens2020} in gray and black dashed lines. We show theoretical predictions from \citet{Liang2019} in the orange dashed line.  }
    \label{fig:tdust_vs_z}
\end{figure}

\subsubsection{$T_{\rm dust}$ versus redshift}

In Figure~\ref{fig:tdust_vs_z}, we show the dust temperature as a function of redshift. The TNG predictions are the $T_{\rm peak}$ and $T_{\rm sed}$ of galaxies with $10^{10}\msun < M_{\ast} < 10^{11}\msun$ in all three simulations. The mass range roughly corresponds to galaxies with $L_{\rm IR} \sim 10^{12}\Lsun$ and is chosen to match the typical value of observed samples at high redshift~\citep[e.g.,][]{Schreiber2018,Bouwens2020,Faisst2020}. Results derived purely from TNG50 galaxies at $z=4$ are also shown and they do not differ considerably from the results of all TNG galaxies. For comparison, we show observational constraints at low redshift from \citet{Schreiber2018} (and the results based on \citet{Bethermin2015} samples), SPT-selected galaxies at $z=2\operatorname{-}6$ in \citet{Strandet2016} \citep[compiled by][]{Bouwens2020}. In addition to these, we include constraints based on the ALPLINE-ALMA galaxies in \citet{Bethermin2020}, the Lyman break galaxy in \citet{Pavesi2016}, the four main-sequence ALMA galaxies in \citet{Faisst2020} (having $T_{\rm sed}$ determined by a $\beta$ free fit but also having $T_{\rm peak}$ measured), $z\gtrsim 6$ galaxies from \citet{Knudsen2016,Hashimoto2019,Bakx2020} re-measured by \citet{Faisst2020} and the $z\sim 7$ galaxy in \citet{Hashimoto2019} with their original temperature measurement. Among these observational constraints, only a set of measurements in \citet{Faisst2020} can be clearly classified as peak dust temperatures (they are marked as triangles in the Figure). Other observations all involve some level of SED fitting with various assumptions on the emissivity index $\beta$ and fitting functions, which could result in model-dependent uncertainties. It is also worth noting that only \citet{Faisst2020} assumed the same multi-component SED model as ours with a general description of emission opacity. Therefore, the $T_{\rm sed}$ we obtain here forms an apple-to-apple comparison only with their results. When compared to other observational studies listed here, one should be cautious in interpreting the differences of the fitted dust temperatures, since results can be dramatically affected by the assumptions of the SED model \citep[see the tests in][for example]{Casey2012}. For reference, we include the models proposed by \citet{Schreiber2018} and \citet{Bouwens2020} based on observational constraints and the $T_{\rm peak}$-$z$ relation from the FIRE-2 simulations~\citep{Liang2019}. 

\begin{figure*}
    \centering
    \includegraphics[width=0.49\textwidth]{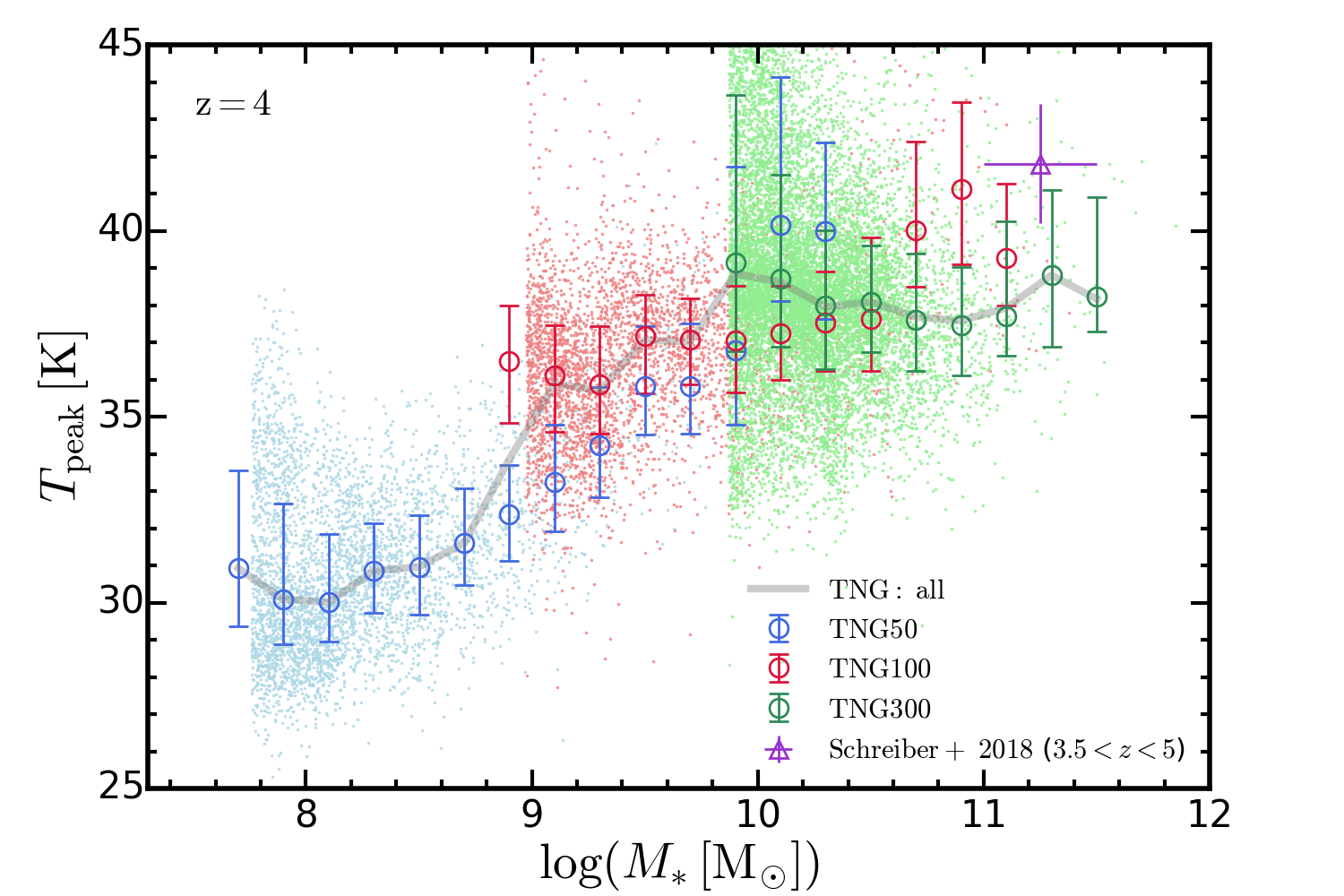}
    \includegraphics[width=0.49\textwidth]{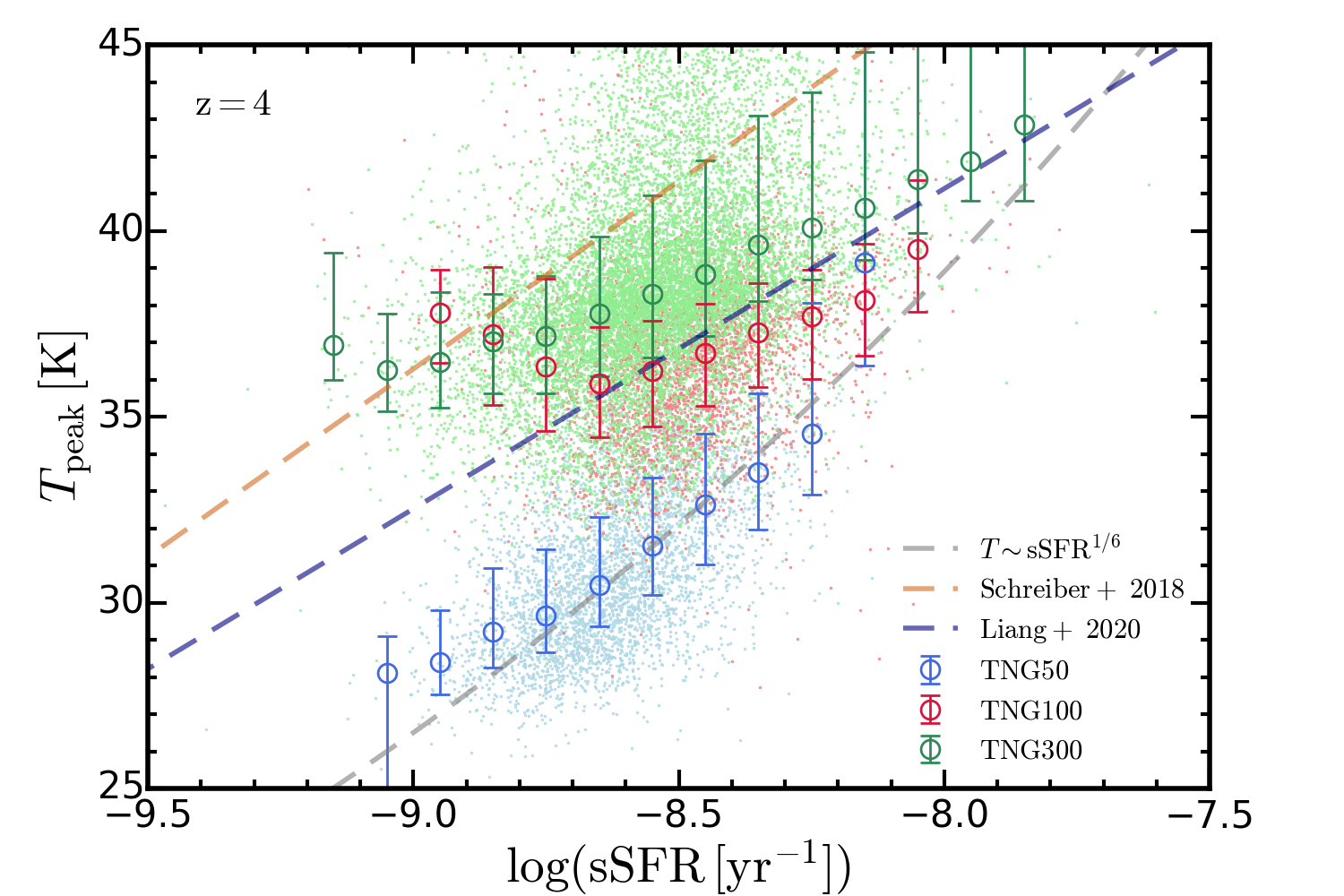}
    \caption{\textbf{Scaling relation of dust temperature.} {\it Left:} Dust temperature versus galaxy stellar mass. We show the galaxies from TNG50/100/300 with $M_{\ast}\geq 1000m_{\rm b}$. The median and $1.5\sigma$ dispersion are shown in circles with error bars for each simulation. The median relation derived based on all TNG galaxies is shown with the gray dashed line. The peak dust temperature shows a positive dependence on galaxy stellar mass at $M_{\ast} \lesssim 10^{10}\msun$ while reaches a plateau at larger masses. The temperature at the massive end is consistent with observational constraints, including the measurement from stacked galaxy SEDs \citep{Schreiber2018} shown here and other measurements at $z\simeq 4$ shown in Figure~\ref{fig:tdust_vs_z}. {\it Right:} Dust temperature versus galaxy specific star formation rate. The notation is the same as the left panel. The gray dashed line shows the relation $T\sim {\rm sSFR}^{1/(4+\beta)}$ ($\beta=2$) for reference. The TNG50 galaxies and the envelope of all TNG galaxies follow the reference relation. Galaxies in TNG100/300 have systematically hotter dust than TNG50 at fixed sSFR, due to the mass dependence of dust temperature shown in the left panel. The results are also compared to theoretical predictions from the FIRE simulations \citep{Liang2020} and the observational study \citep{Schreiber2018}.
    }
    \label{fig:tdust_vs_mass_sSFR}
\end{figure*}

In general, the dust temperatures in both observations and simulations are higher at high redshift than in the local Universe. This can be understood by the correlation of dust temperature and the specific star formation rate~\citep[see also][]{Magnelli2014,Ma2019}
\begin{align}
T_{\rm dust} & \sim \Big(\dfrac{L_{\rm IR}}{M_{\rm dust}}\Big)^{1/(4+\beta)} \sim \Big(\dfrac{\rm SFR}{M_{\rm dust}}\Big)^{1/(4+\beta)} \nonumber \\
& \sim \Big(\dfrac{\rm SFR}{M_{\ast}} \dfrac{M_{\ast}}{M_{\rm metal}} \dfrac{M_{\rm metal}}{M_{\rm dust}}\Big)^{1/(4+\beta)} \nonumber \\
& \sim \Big[{\rm sSFR} \dfrac{1}{M_{\rm metal}/M_{\ast}} \Big(\dfrac{1}{\rm DTM}-1 \Big)\Big]^{1/(4+\beta)},
\label{eq:T-ssfr}
\end{align} 
where sSFR is the specific star formation rate and DTM is the dust-to-metal ratio. We have used $L_{\rm IR}\sim {\rm SFR}$ and ${\rm DTM} \equiv M_{\rm dust}/(M_{\rm dust} + M_{\rm metal})$ in the derivation above. The average sSFR increases at high redshift and reaches a plateau at $z\sim 4$~\citep[e.g.,][]{Tomczak2016,Santini2017}. In addition, the metal-to-stellar mass ratio decreases at high redshift owing to the shift to lower mass on the mass-metallicity relation and the potentially decreasing normalization of the mass-metallicity relation at high redshift~\citep{Ma2016}. Moreover, the DTM could also be lower at high redshift, as found in the calibration procedure in \citetalias{Vogelsberger2020} and other studies~\citep[e.g.,][]{Inoue2003,McKinnon2016,Aoyama2017,Behrens2018}. These factors will drive the dust temperature to be higher at high redshift, which is consistent with the phenomena shown in Figure~\ref{fig:tdust_vs_lir}. Compared to observations, the TNG predicted $T_{\rm peak}$ at $z=4$ is consistent with observations while the $T_{\rm sed}$ from TNG is significantly higher than both observational constraints and $T_{\rm peak}$. When optically thick dust self-absorption is considered in SED fitting, it is actually expected that $T_{\rm sed}$ will be considerably larger than $T_{\rm peak}$. For example, $T_{\rm peak} \simeq 40\,{\rm K}$ could correspond to $T_{\rm sed}\simeq 60\,{\rm K}$ as shown in \citet{Casey2012b} and \citet{Liang2019}. At $z\geq 6$, the SED dust temperature from TNG is still about $20\,{\rm K}$ higher than the temperature measured in observations. At this redshift range, the peak dust temperature from TNG follows the models in \citet{Schreiber2018} and \citet{Bouwens2020}, which linearly rise up at higher redshift. However, such high $T_{\rm peak}$ is in tension with the peak dust temperature measurements of $z\gtrsim 5$ galaxies compiled in \citet{Faisst2020}, which suggests a plateau of $T_{\rm peak}$ at $z\gtrsim 4$. This behavior is in agreement with the FIRE-2 predictions which are based on simulations of higher spatial and mass resolution. In conclusion, these comparisons here suggest that the TNG predicted dust temperatures are higher than observations at $z\geq 6$. 

\subsubsection{$T_{\rm dust}$ versus $M_{\ast}$ and sSFR}

In the left panel of Figure~\ref{fig:tdust_vs_mass_sSFR}, we show the peak dust temperature as a function of galaxy stellar mass at $z=4$. The binned estimations derived from TNG50/100/300 are shown explicitly along with scatter plot of individual galaxies. For each simulation, we only select galaxies with $M_{\ast}>1000m_{\rm b}$ for analysis. In general, the dust temperature increases with galaxy stellar mass when $M_{\ast}\lesssim 10^{10}\msun$ and reaches a plateau when $M_{\ast}\gtrsim 10^{10}\msun$. The plateau feature is similar to what we found in the $T_{\rm peak}$-$L_{\rm IR}$ relation and is consistent with the weak dependence of $T_{\rm peak}$ on $M_{\ast}$ of $M_{\ast}\gtrsim 10^{9}\msun$ galaxies found in \citet{Liang2020}. For massive galaxies, the predicted dust temperature is roughly consistent with the stacked SED measurement of stellar mass-binned samples in \citet{Schreiber2018} and other measurements at $z\simeq 4$ shown in Figure~\ref{fig:tdust_vs_z}. Results from different simulations are consistent with each other in their shared dynamical ranges, but simulation with poorer resolution always exhibits larger scatter and slight shift towards higher temperature. This marks the potential overprediction of dust temperature in galaxies with poor sampling of radiation sources (and dust attenuating gas cells as well). In the right panel of Figure~\ref{fig:tdust_vs_mass_sSFR}, we show the peak dust temperature as a function of galaxy sSFR. For reference, the gray dashed line shows the relation $T\sim {\rm sSFR}^{1/(4+\beta)}$ ($\beta=2$) as in Equation~\ref{eq:T-ssfr}. Although TNG galaxies show significant dispersion in this plane, the low temperature envelope of the distribution roughly follows the reference relation. TNG50 galaxies follow the reference relation while TNG100/300 galaxies have systematically hotter dust component at fixed sSFR. This is related to the mass dependence of dust temperature shown in the left panel. We compare our results with relations reported in the observational \citep{Schreiber2018} and theoretical studies \citep{Liang2020}. \citet{Liang2020} found the correlation between $\Delta T_{\rm dust} \equiv T_{\rm dust} - T^{\rm med}_{\rm dust}$ and the star formation burstiness $SB \equiv {\rm sSFR} - {\rm sSFR}^{\rm med}$, where ``med'' indicates the median value of their sample. We convert the relation to the $T_{\rm dust}$-sSFR plane using the reported median values therein. Similarly, \citet{Schreiber2018} found the $\Delta T_{\rm dust}$-SB correlation but the reference points are the main sequence dust temperature and sSFR, for which we use the $T^{\rm MS}_{\rm dust}(z)$ therein and $\log{{\rm sSFR}^{\rm MS}}(z=4)\simeq -8.5$, which is consistent with observational constraints \citep[e.g.,][]{Salmon2015,Tomczak2016,Santini2017} as well as TNG galaxies shown here. All of the sampled galaxies in \citet{Liang2020} at $z=4$ have $M_{\ast}\gtrsim 10^{9}\msun$. As expected, their relation is quite consistent with TNG100/300 galaxies with comparable stellar masses. The \citet{Schreiber2018} samples are even more massive galaxies ($M_{\ast}\gtrsim 10^{10}\msun$). Their relation lies slightly above the TNG300 predictions and is consistent with the same level of difference in Figure~\ref{fig:tdust_vs_z}. 

\subsubsection{Discussions}

Equation~\ref{eq:T-ssfr} indicates that the dust temperature is related to the dust mass when $L_{\rm IR}$ is fixed. Since most of the theoretical works have dust models calibrated based on UV luminosities of galaxies, the total amount of energy absorbed and re-emitted at IR wavelengths should match observations by construction, regardless of the dust model adopted. However, the dust temperature offers another channel to test theoretical predictions since it anti-correlates with the actual amount of dust predicted when $L^{\rm abs}_{\rm UV} \sim L_{\rm IR}$ is fixed. Here, TNG predicts higher peak dust temperatures than some observations at $z\geq 6$ as well as results from simulations of higher resolution indicating that the dust mass could be underpredicted. In \citet{Vogelsberger2020}, we found that the average dust-to-metal (DTM) ratio follows a $(z/2)^{-1.92}$ law at $z\geq 2$, which implies ${\rm DTM}\lesssim 0.1$ at $z\gtrsim 6$ much lower than the value in the local Universe $\sim 0.4$~\citep{Dwek1998}. Such a low predicted dust abundance could be attributed to the limited spatial and mass resolution of high-redshift galaxies in the TNG simulations. As discussed in \citet{Cen2014}, if the porosity (and clumpiness) of the ISM is not well-resolved, the dust opacity of radiation sources can be overestimated. In the calibration procedure, less dust will be required to produce the same amount of attenuation. \citet{Cen2014} has artificially allowed a fraction $f_{\rm esc}$ of the intrinsic radiation leak from their $z\sim 7$ galaxies to correct for this effect and the best-fit DTM rose to $0.4$ with $f_{\rm esc}=0.1$, as opposed to the DTM of $0.06$ they found originally. Another potential caveat of the dust model is that the DTM is set to a constant for all the galaxies at a given redshift. The dust abundance of a small sub-population of massive star-forming galaxies, which are completely dark in UV/NIR searches, could be heavily under-estimated in this calibration process. This could serve as a potential explanation to the missing IR bright galaxies in TNG without involving the galaxy formation physics models of the simulations.

In the meantime, if the dust-attenuating medium is not well-resolved, the optical depth of the hot dust component heated at the vicinity of radiation sources could be underestimated (i.e. a single layer of gas cells covering a radiation source are heated while their dust emission are not shielded at all), which will result in higher light-weighted dust temperatures as well. Such effect appears in the left panel of Figure~\ref{fig:tdust_vs_mass_sSFR}. At similar stellar mass, the result from low-resolution simulation exhibits larger scatter and systematic shift towards higher temperature. For each simulation, the scatter is also larger at smaller stellar mass and, in a small fraction of poorly resolved galaxies (at the edge of our stellar mass cut), dust can be heated to $\gtrsim 3\sigma$ higher than the median temperature. 

The measurements or assumptions of dust temperature could also affect the estimation of the bolometric IR luminosity. For example, if the observational data only cover the Rayleigh--Jeans tail (long wavelengths) of the IR SED, the inferred bolometric luminosity will be sensitive to the dust temperature assumed. Given the same flux measurement at long wavelengths, the higher the dust temperature assumed, the higher the bolometric luminosity will be determined. However, in terms of the discrepancy we found for the IR luminosity function in Section~\ref{sec:results-irlf}, this effect would actually exaggerate the mismatch at the bright end, since the observational studies typically assume lower dust temperatures than the simulation results (i.e. increasing the dust temperature will lead to even higher inferred bolometric IR luminosities and make it even more challenging for the TNG simulations to match). 

\section{Conclusions}
\label{sec:conclusions}

In this paper, we have expanded the theoretical predictions of high-redshift galaxies from IllustrisTNG to IR wavelengths, using {\sc Skirt} radiative transfer calculations. The analysis pipeline has been adapted from \citetalias{Vogelsberger2020} to self-consistently model dust emission and self-absorption at IR wavelengths. The pipeline produces a catalog of high-redshift galaxies with their NIR-to-FIR SEDs calculated in this paper along with their UV-optical high resolution SEDs calculated in \citetalias{Vogelsberger2020}. Based on this, we make various NIR and FIR predictions of galaxy populations at $z\geq 4$. Our findings can be summarized as follows:
\begin{itemize}
    \item The predicted rest-frame $K$-band and $z$-band luminosity functions at $z\geq 4$ are presented and compared with existing observations. The predictions are consistent with observations, despite a slight underprediction at the bright end of the $z=4$ $K$-band and $z$-band luminosity functions. The luminosity functions in both bands are tracers of galaxy stellar mass assembly. The faint-end luminosity functions evolve in a self-similar way at $z\geq 4$ with a roughly constant faint-end slope. This indicates that the star-forming galaxies at this epoch have roughly the same mass doubling time. 
    
    \item Assuming $100\%$ survey completeness, we make theoretical predictions for the {\it JWST} MIRI apparent band luminosity functions and number counts. At $z=6$, $\sim 3000$ ($\sim 500$) galaxies are expected in the  F560W (F1000W) band assuming a survey area of $500\,{\rm arcmin}^2$ and depth of $\Delta z=1$. Large NIR galaxy surveys conducted with MIRI can be about $2\mmag$ deeper than the current best observations.

    \item We make predictions for the bolometric IR luminosity functions at $z\geq 4$. The results are mainly affected by the predicted FIR dust continuum emission. A better agreement with observations is achieved at the faint end compared to previous theoretical attempts. However, the abundance of most luminous IR galaxies ($L_{\rm IR} \sim 10^{13} \Lsun$) is significantly underpredicted ($\sim 1\,\mathrm{dex}$ deficiency) by TNG. The discrepancy consistently shows up at $z=4$ and $z=6$, and is potentially related with the underpredicted counts of sub-millimeter galaxies reported in previous works based on TNG~\citep{Hayward2020}. The tension cannot be resolved by considering merging systems in TNG.
    
    \item By integrating over the bolometric IR luminosity function, we make predictions for the obscured cosmic SFRD at $z\geq 4$. Combining the rest-frame UV luminosity function in \citetalias{Vogelsberger2020} and the ${\rm H}\alpha$ luminosity function in \citetalias{Shen2020}, we are able to compare the cosmic SFRD traced by different indicators. The predicted unobscured SFRD (traced by UV and ${\rm H}\alpha$) is consistent with observations. The total SFRD derived by summing up ${\rm SFRD}_{\rm UV}$ and ${\rm SFRD}_{\rm IR}$ agrees remarkably well with the instantaneous SFRD traced by gas cells in simulations. The obscured SFRD predicted from TNG suggests that it becomes subdominant (contributes less than $50\%$ to the total SFRD) at $z\gtrsim 5$ and diminishes at higher redshift. Such a prediction is in tension with the recent ALPINE-ALMA survey which suggests a significant contribution of unobscured SFRD at $4\lesssim z \lesssim 6$.

    \item Specifically, we check the SFRD contributed by the most obscured and thus most luminous IR galaxies ($L_{\rm IR} \gtrsim 10^{12} \Lsun$) in simulations and compare it to the SFRD of optical/NIR dark galaxies revealed in IR/sub-millimeter observations. The SFRD in such galaxies in simulations is about $1\sigma$ lower than the results from the ALPINE-ALMA survey.
    
    \item We make predictions for the dust temperature of high-redshift galaxies. We find that the dust temperature positively correlates with both galaxy bolometric IR luminosity and redshift. The predicted peak dust temperature in typical IR galaxies is about $40\,{\rm K}$ ($60\,{\rm K}$) at $z=4$ ($z=6$). The SED dust temperature is systematically higher than $T_{\rm peak}$ by about $20\,{\rm K}$. The predicted $T_{\rm peak}$-$L_{\rm IR}$ relation is shifted from the local relation to higher temperatures while remaining consistent with the sub-millimeter galaxies selected at $z \simeq 4$. However, at $z\geq 6$, we find that TNG overpredicts the peak dust temperature of galaxies by about $20\,{\rm K}$. This overprediction of dust temperature could be related to the low dust-to-metal ratio of our model, and could be attributed to the limitation of the simulations in resolving the porosity and clumpiness of the ISM. 
    
\end{itemize}

In conclusion, similar to the previous comparisons in rest-frame UV/optical in \citetalias{Vogelsberger2020} and \citetalias{Shen2020}, the NIR properties of high-redshift galaxies in TNG are quite consistent with observations. Predictions for the {\it JWST} MIRI instrument are calculated, which leads to a complete {\it JWST} galaxy multiband photometric catalog when combined with the NIRCam predictions in \citetalias{Vogelsberger2020}. However, we find large, systematic discrepancies of FIR properties of most luminous IR galaxies in TNG compared to recent observations. The abundance of luminous IR galaxies and their contribution to the cosmic SFRD are underpredicted by TNG. The solution to this would require both higher intrinsic on-going star formation and stronger metal and dust enrichment in high-redshift galaxies. The discrepancy found here could serve as a constraint on the sub-grid feedback model of cosmological simulations.

%%%%%%%%%%%%%%%%%%%%%%%%%%%%%%%%%%%%%%%%%%%%%%%%%%

\section*{Acknowledgements}
MV acknowledges support through NASA ATP grants 16-ATP16-0167, 19-ATP19-0019, 19-ATP19-0020, 19-ATP19-0167, and NSF grants AST-1814053, AST-1814259, AST-1909831 and AST-2007355. 
ST is supported by the Smithsonian Astrophysical Observatory through the CfA Fellowship. 
PT acknowledges support from NSF grant AST-2008490. 
FM acknowledges support through the Program ``Rita Levi Montalcini'' of the Italian MUR.
The primary TNG simulations were realized with compute time granted by the Gauss Centre for Super-computing (GCS): TNG50 under GCS Large-Scale Project GCS-DWAR (2016; PIs Nelson/Pillepich), and TNG100 and TNG300 under GCS-ILLU (2014; PI Springel) on the GCS share of the supercomputer Hazel Hen at the High Performance Computing Center Stuttgart (HLRS). 

\section*{Data Availability}
The simulation data of the IllustrisTNG project is publicly available at \href{https://www.tng-project.org/data/}{https://www.tng-project.org/data/}. The analysis data of this work was generated and stored on the super-computing system Cannon at Harvard University. The data underlying this article can be shared on reasonable request to the corresponding author.



%\bibliographystyle{mnras}
%\bibliography{reference} % if your bibtex file is called example.bib

%%%%%%%%%%%%%%%%%%%%%%%%%%%%%%%%%%%%%%%%%%%%%%%%%%

%%%%%%%%%%%%%%%%% APPENDICES %%%%%%%%%%%%%%%%%%%%%

\appendix
\section{Test with different {\sc Skirt} configurations}
\label{appsec:skirt_test}

Compared to the procedure in \citetalias{Vogelsberger2020}, the dust model and several parameters in the {\sc Skirt} calculations have been modified for the predictions in the IR. In this section, we will test how these modifications impact the resulting galaxy SED and discuss our parameter choices.

\begin{figure}
    \centering
    \includegraphics[width=0.49\textwidth]{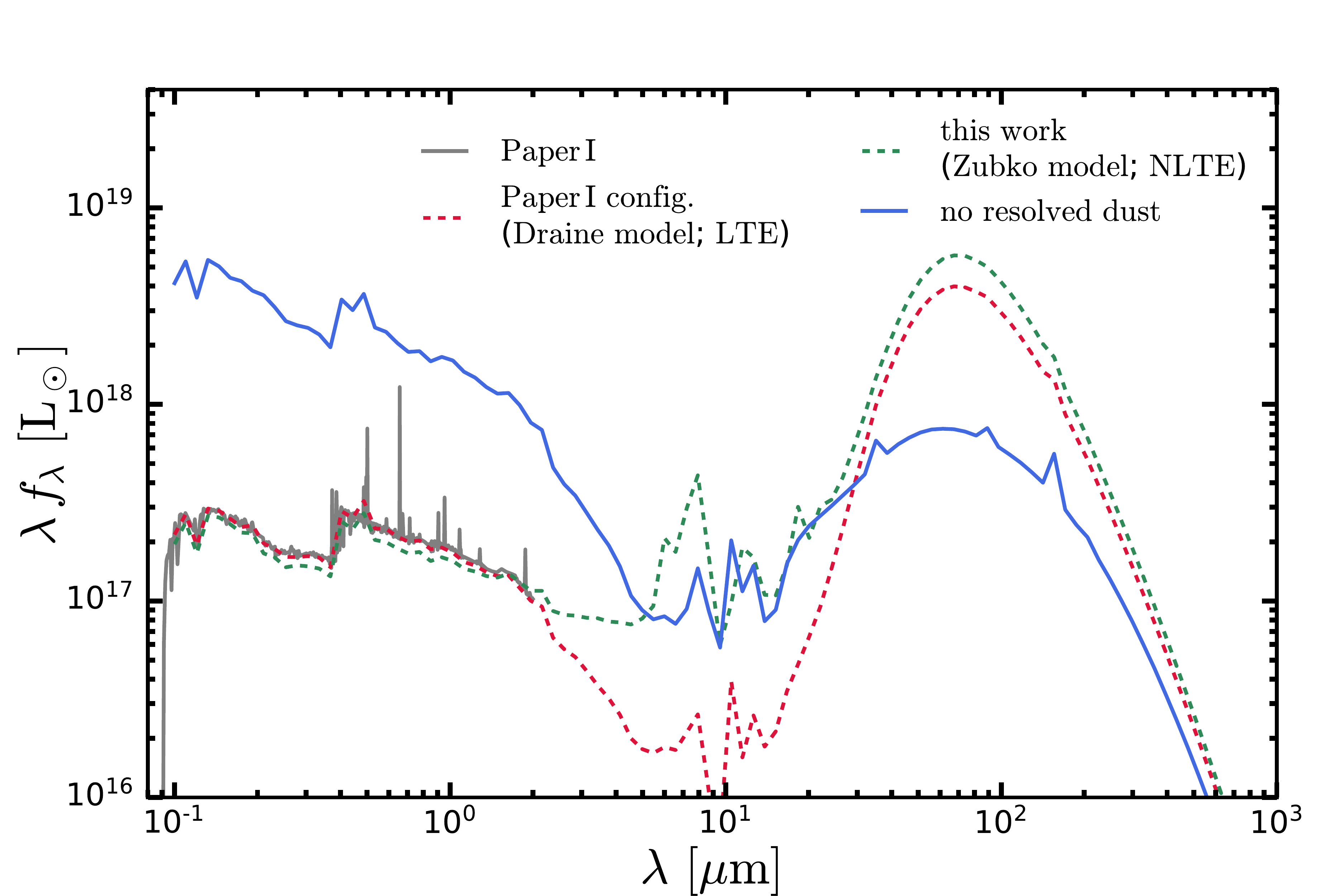}
    \includegraphics[width=0.49\textwidth]{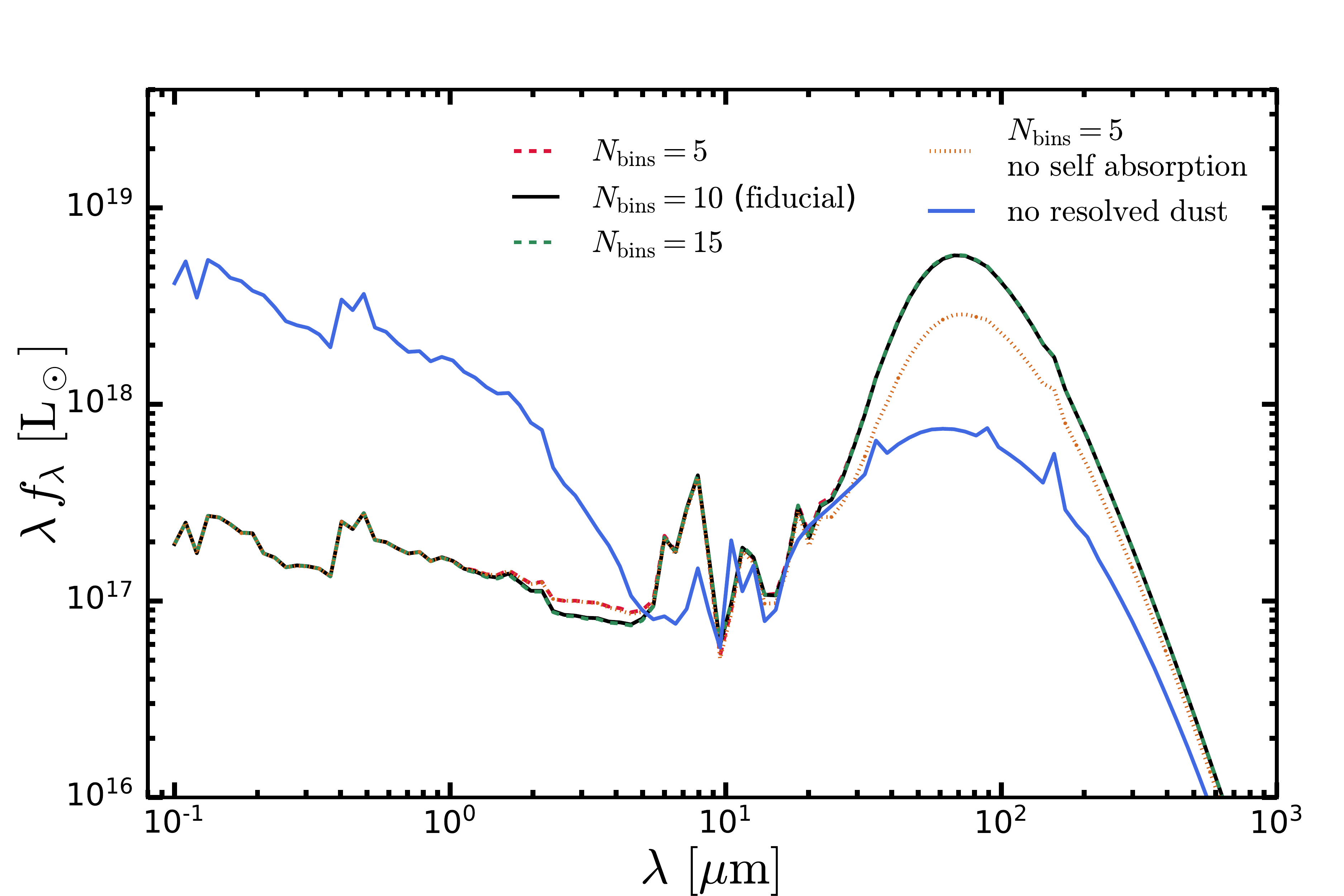}
    \caption{\textbf{Galaxy SED calculated by different {\sc Skirt} configurations.} {\it Top:} We show the SEDs of a well-resolved galaxy (with $\sim 1.4\times 10^{5}$ stellar particles) at $z=4$ in TNG100. The UV-to-NIR SED calculated in \citetalias{Vogelsberger2020} is shown in the gray solid line. The SED without resolved dust attenuation is shown in blue as a reference. The SED (that extends to IR wavelengths) calculated with the original dust model in \citetalias{Vogelsberger2020}, which takes the average dust properties of the \citet{Draine2007} dust mixture, is shown in red. The SED calculated with the \citet{Zubko2004} multi-grain dust model adopted in this work, including the non-local thermal equilibrium of small grains, is shown in green. The major difference between the new model and the old one is the enhanced flux at mid-IR and FIR wavelengths. The UV-to-optical attenuation has also been enhanced due to the model switch, which has been compensated by tuning the dust-to-metal ratio as discussed in the main text. {\it Bottom:} We show the SEDs of the same galaxy as the top panel but with variations in {\sc Skirt} configurations from the fiducial setup. The fiducial results with number of grain size bins $N_{\rm bin}=10$ is shown in black. The SED calculated with $N_{\rm bin}=5$ ($N_{\rm bin}=15$) is shown in red (green). The comparison demonstrates the convergence of the SED when $N_{\rm bin}\geq 10$. In addition, we show the SED calculated without dust self-absorption and the major change induced by the self-absorption is the enhanced flux at the FIR peak.
    }
    \label{appfig:sed_skirt_config}
\end{figure}

The first difference in the {\sc Skirt} setup is that we have switched to the \citet{Zubko2004} multi-grain dust mixture (from the average dust property of the \citet{Draine2007} dust mixture), in order to trace grains of small sizes independently and consider their decoupling from local thermal equilibrium. In the top panel of Figure~\ref{appfig:sed_skirt_config}, we show that the main impact of this switch on galaxy SEDs is the enhancement of flux in the mid-IR and FIR. Compared to the UV-to-optical SED produced in \citetalias{Vogelsberger2020}, the attenuation in the UV is slightly stronger, so we decide to decrease the dust-to-metal ratio to compensate this effect and maintain consistency of the UV predictions with observations. 
In the bottom panel of Figure~\ref{appfig:sed_skirt_config}, we show SEDs calculated if we vary the setup (wavelength gird, dust model, dust self-absorption, number of bins of grain sizes) from the fiducial one introduced in the main text (see Section~\ref{sec:method}). The number of bins for dust grain sizes has impact on the mid-IR SED but the results converge when $N_{\rm bins}\geq 10$. The dust self-absorption is important for the prediction at the FIR peak and the flux there can be enhanced by $\sim 0.3\,\mathrm{dex}$ compared to the one without self-absorption.

\begin{figure}
    \centering
    \includegraphics[width=0.49\textwidth]{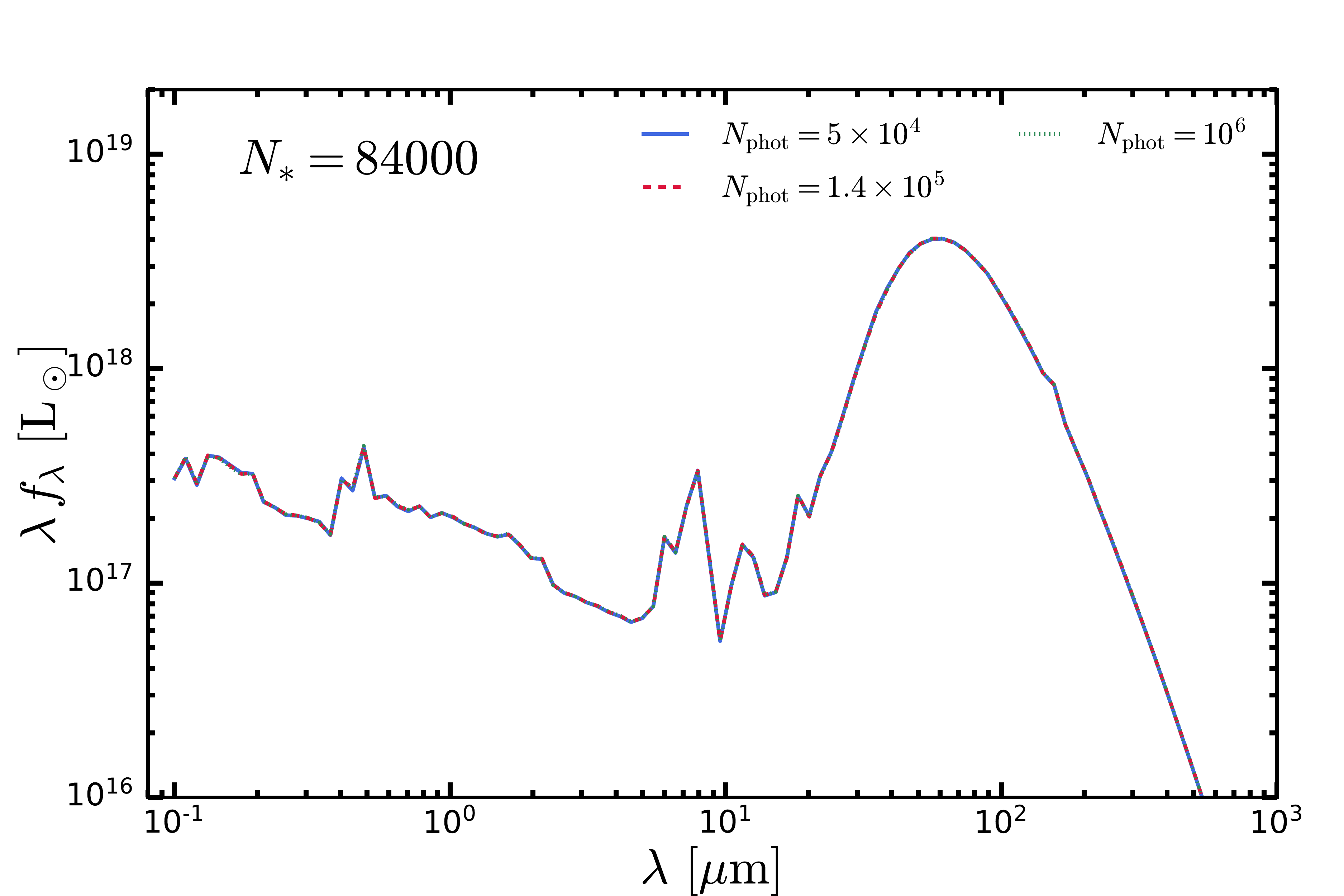}
    \includegraphics[width=0.49\textwidth]{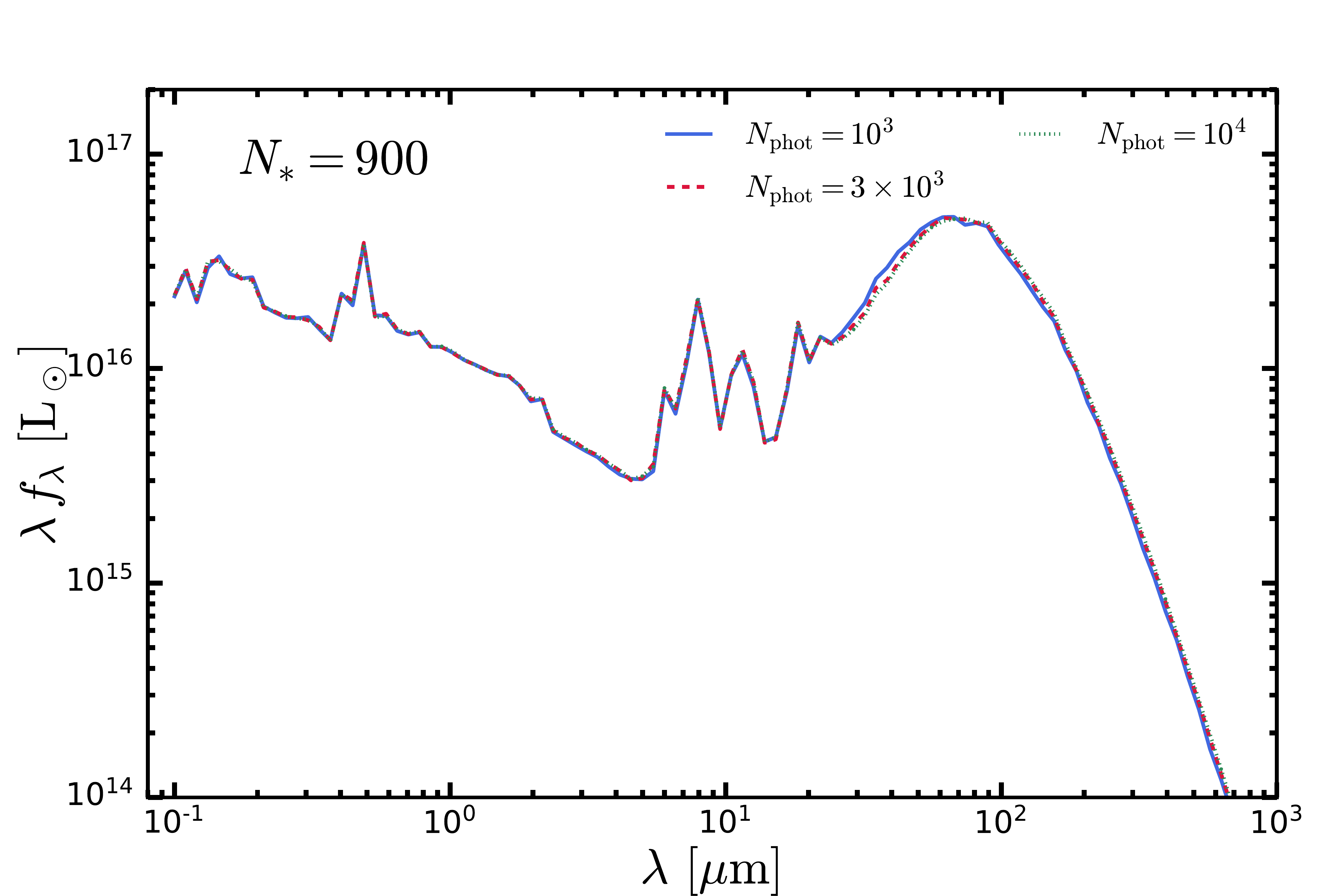}
    \includegraphics[width=0.49\textwidth]{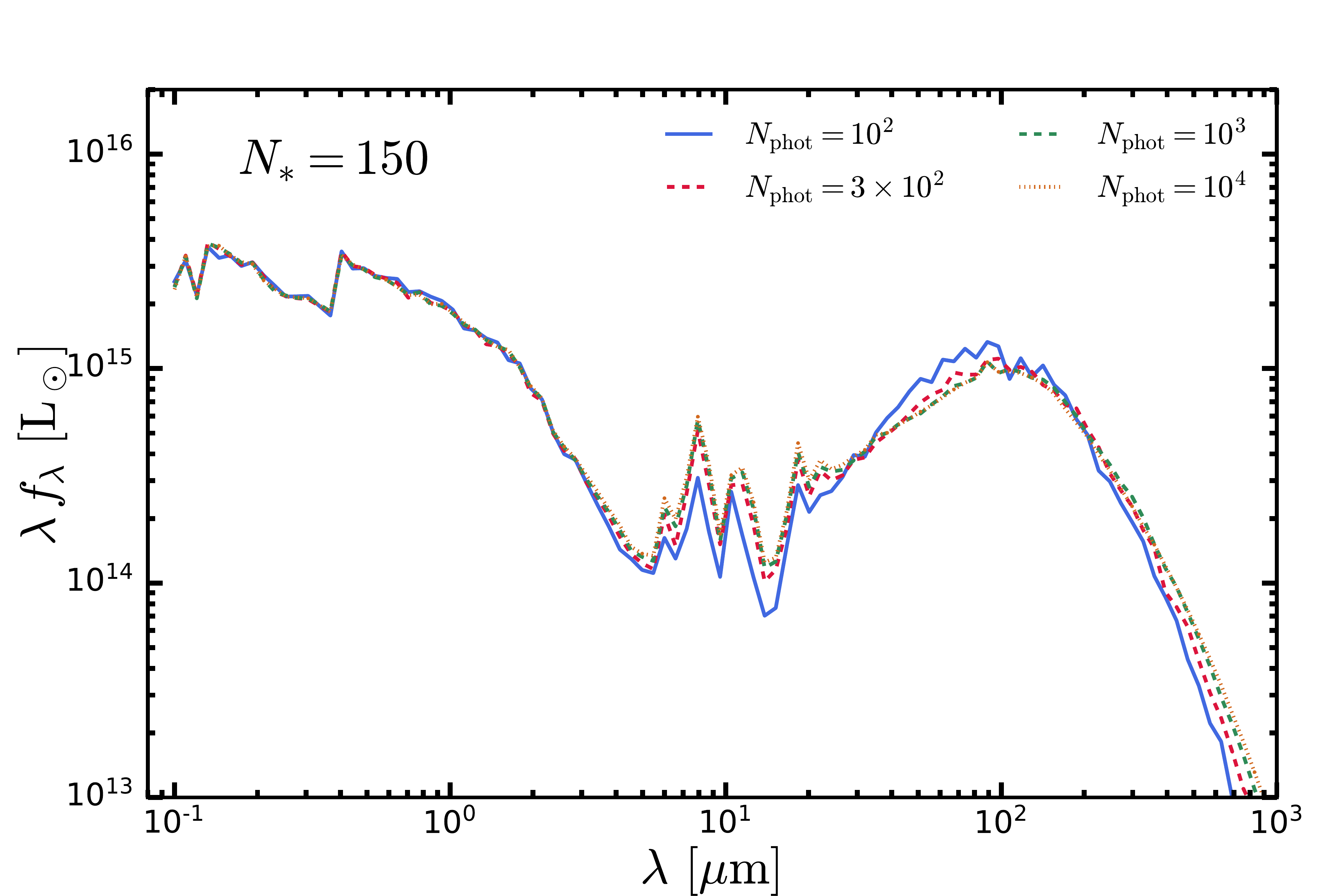}
    \caption{\textbf{Galaxy SEDs calculated with different number of photon packages.} For presentation, we pick three galaxies selected at $z=4$ in TNG100 as examples. {\it Top:} A well-resolved galaxy with $N_{\ast}=84000$ in the input of {\sc Skirt} calculation. The SED is converged with $N_{\rm phot}=50000 < N_{\ast}$. {\it Middle:} A resolved galaxy with $N_{\ast}=900$ and the SED is converged with $N_{\rm phot}=1000 \sim N_{\ast}$, despite tiny differences at $\lambda \sim 40\micron$. {\it Bottom:} A poorly-resolved galaxy with $N_{\ast}=150$. The SED with $N_{\rm phot}=100$ underpredicts the flux in the mid-IR and overpredicts the flux at the rising edge of the FIR peak (by $\lesssim 0.1\,\mathrm{dex}$), but the SED is converged with $N_{\rm phot}\geq 300 \sim 2 \times N_{\ast}$.
    }
    \label{appfig:sed_nphot}
\end{figure}

In addition, due to various changes in the {\sc Skirt} setup (see Section~\ref{sec:method}), the requirement on the number of photon packages, $N_{\rm phot}$, to reach convergence may vary from the one in \citetalias{Vogelsberger2020}. To test this, we choose several galaxies with different numbers of stellar particles from TNG100 and perform {\sc Skirt} calculations with different $N_{\rm phot}$. In Figure~\ref{appfig:sed_nphot}, we show the results of three galaxies as examples and we note that the behaviour of all tested galaxies is similar. For the well-resolved galaxy with $N_{\ast}=84000$, the SED is converged with $N_{\rm phot}=50000$, which is smaller than $N_{\ast}$. For the galaxy with $N_{\ast}=900$, the SED is also converged with $N_{\rm phot} \gtrsim N_{\ast}$, despite tiny differences at $\lambda \sim 40\micron$. For the poorly-resolved galaxy with $N_{\ast}=150$, the SED is converged with $N_{\rm phot} \gtrsim 2\times N_{\ast}$ while having a steeper rising edge of the FIR peak when $N_{\rm phot}=100$. As discussed in the main text, we set $N_{\rm phot} = N_{\ast}$ for radiative transfer calculations of galaxies in TNG100 and TNG300. For low mass, poorly-resolved galaxies in these two simulations, the uncertainties from the {\sc Skirt} calculations and from the unresolved physical processes of the simulations are degenerate, but these uncertainties can be captured to first order by the resolution correction procedure with TNG50 as a reference. Meanwhile, we choose $N_{\rm phot} = 3\times N_{\ast}$ for TNG50 galaxies to ensure the convergence of the SEDs of poorly-resolved galaxies in this simulation, which does not have a simulation in the TNG suite with higher resolution for correction. The comparisons shown in Figure~\ref{appfig:sed_nphot} justify our parameter choices in this work.

\section{Best-fit Schechter function parameters}
\label{appsec:schfit}

In Table~\ref{apptab:schfits}, we present the best-fit Schechter function parameters for the rest-frame FUV/bolometric IR luminosity functions and the {\it JWST} MIRI band apparent luminosity functions. The fitting functions has been described in Section~\ref{sec:results-jwstlf}. The FUV results are taken from \citetalias{Vogelsberger2020}.

\begin{table}
\caption{{\bf Parameters of the best-fit Schechter functions to the luminosity functions.} The table contains the best-fit faint-end slope $\alpha$, number density normalization $\phi^{\ast}$ and break luminosity $L^{\ast}$/break magnitude $M^{\ast}$ for the rest-frame FUV/bolometric IR luminosity functions and the {\it JWST} MIRI band apparent luminosity functions from the TNG simulations.}
\centering
\begin{tabular}{p{0.13\textwidth}|p{0.05\textwidth}|p{0.035\textwidth}|p{0.05\textwidth}|p{0.1\textwidth}}
\hline 
{\bf Band} & redshift & $\alpha$ & $M^{\ast}$ & $\log{\phi^{\ast}}$  \\
 &  &  & [${\rm mag}$] & [$1/\Mpc^{3}/\mmag$]\\
\hline
\hline
{\bf FUV (rest-frame)}    & 2  & -1.58 & -20.45 & -2.65 \\
                          & 4  & -1.80 & -21.10 & -3.04 \\
                          & 6  & -2.04 & -21.31 & -3.61 \\
                          & 8  & -2.45 & -21.44 & -4.71 \\
\hline
\vspace{0.4cm}\\

\hline 
{\bf Band} & redshift & $\alpha$ & $m^{\ast}$ & $\log{\phi^{\ast}}$  \\
 &  &  & [${\rm mag}$] & [$1/\Mpc^{3}/\mmag$]\\
\hline
\hline
{\bf F560W}  & 4  & -1.91  & 22.50 & -4.01\\
             & 6  & -1.80  & 24.53 & -3.89\\
             & 8  & -2.20  & 25.27 & -5.00\\
\hline
{\bf F1000W} & 4  & -1.77 & 22.50 & -3.75\\
             & 6  & -1.77 & 23.91 & -4.15\\
             & 8  & -2.28 & 24.39 & -5.38\\
\hline
{\bf F1280W} & 4  & -1.73 & 22.61 & -3.71 \\
             & 6  & -1.67 & 23.71 & -4.08 \\
             & 8  & -1.91 & 24.83 & -5.12 \\
\hline
{\bf F1500W} & 4  & -1.70 & 22.30 & -3.77\\
             & 6  & -1.68 & 23.53 & -4.25\\
             & 8  & -2.10 & 24.21 & -5.51\\
\hline  
\vspace{0.4cm}\\

\hline 
{\bf Band} & redshift & $\alpha$ & $\log{L^{\ast}}$ & $\log{\phi^{\ast}}$  \\
 &  &  & [$\Lsun$] & [$1/\Mpc^{3}/{\rm dex}$]\\
\hline
\hline 
{\bf Bolometric IR}      & 4 & -1.73 & 12.25 & -4.35\\
\,\,\,{\bf (rest-frame)} & 6 & -1.61 & 12.08 & -4.70\\
                         & 8 & -1.79 & 12    & -5.95\\
\hline
\end{tabular}
\label{apptab:schfits}
\end{table}

%%%%%%%%%%%%%%%%%%%%%%%%%%%%%%%%%%%%%%%%%%%%%%%%%%

% Don't change these lines
\bsp	% typesetting comment
\label{lastpage}
\end{document}